\documentclass[12pt,preprint]{aastex}

\newcommand{\gf}{{\it gf}}
\newcommand{\vt}{v$_{\rm t}$}
\newcommand{\teff}{T$_{\rm eff}$}
\newcommand{\logg}{log~$g$}
\newcommand{\msun}{M$_{\odot}$}
\newcommand{\tphot}{T$_{\rm Phot}$}
\newcommand{\tion}{T$_{\rm Ion}$}
\newcommand{\tex}{T$_{\rm Ex}$}
\newcommand{\leps}{$\log{\epsilon}$}

\shorttitle{Abundances in Baade's Window}
\shortauthors{Fulbright et al.}

\begin{document}

\title{Abundances of Baade's Window Giants from Keck/HIRES Spectra:  I.  Stellar Parameters and [Fe/H] Values\footnote{Based on data obtained at the W. M. Keck Observatory, which is operated as a scientific partnership among the California Institute of Technology, the University of California, and NASA, and was made possible by the generous financial support of the W. M. Keck Foundation.}}

\author{Jon. P. Fulbright}
\affil{Observatories of the Carnegie Institution of Washington, 813 Santa Barbara St., Pasadena, CA 91101}
\email{jfulb@ociw.edu}

\author{Andrew McWilliam}
\affil{Observatories of the Carnegie Institution of Washington, 813 Santa Barbara St., Pasadena, CA 91101}
\email{andy@ociw.edu}

\author{R. Michael Rich}
\affil{Division of Astronomy, Department of Physics and Astronomy, UCLA, Los Angeles, CA  90095-1562}
\email{rmr@astro.ucla.edu}

\begin{abstract}
We present the first results of a new abundance survey of the Milky Way bulge 
based on Keck/HIRES spectra of 27 K-giants in the Baade's Window ($l = 1$, 
$b = -4$) field.  The spectral data used in this study are of much higher 
resolution and signal-to-noise than previous optical studies of Galactic bulge 
stars.  The [Fe/H] values of our stars, which range between $-1.29$ and 
$+0.51$, were used to recalibrate large low resolution surveys of bulge
stars.  Our best value for the mean [Fe/H] of the bulge is $-0.10 \pm 0.04$.
This mean value is similar to the mean metallicity of the local disk and 
indicates that there cannot be a strong metallicity gradient inside the solar
circle.  The metallicity distribution of stars confirms that the bulge 
does not suffer from the so-called ``G-dwarf'' problem.
This paper also details the new abundance techniques necessary
to analyze very metal-rich K-giants, including a new Fe line list and
regions of low blanketing for continuum identification.  

\end{abstract}


\section{Introduction}

We have been engaged in a long term program using high-resolution spectroscopy
to study the composition of stars in the Galactic bulge, with
the aim of constraining the conditions of the bulge's formation
and chemical evolution at a level of detail that is impossible
to obtain with any other method.  Our plan is to obtain high
S/N spectra for samples of $>$20 stars in
each of 3 latitudes in the bulge:  Baade's Window ($b$=$-$4$^{\circ}$),
Sgr~I ($b$=$-$2.65$^{\circ}$), and the Plaut field ($b$=$-$8$^{\circ}$).
This paper describes the beginning of the analysis of the 
Baade's Window sample based on 
Keck/HIRES data; the observations and analysis of the other two fields, taken 
with Magellan/MIKE, will be described in future papers.

The full sample will enable the homogeneity of the
chemical properties of the bulge to be investigated, which may reveal
evidence of galactic accretion, or multiple star formation events.
Pertinent questions might include: What are the metallicity and composition
gradients in the Galactic bulge, and how do they constrain bulge formation
scenarios.  Did some fraction of the population form in separate systems that
merged very early on?  What fraction of the population might have originated
in a system similar to the Sagittarius dwarf spheroidal galaxy?  Are there
correlations between kinematics and composition, and do the composition 
gradients differ from the iron abundance gradient?

The fossil record of the Local
Group is important because we cannot trace, with certainty, the evolution
of distant galaxies or populations in distant galaxies to the present epoch.
In order to understand the formation of galaxies, such as our own, we
must also use the constraints available from the ages, kinematics, and
chemistry of stars in its constituent stellar populations.
This is especially true if present day galaxies
accumulated through a complicated set of mergers, as suggested in
the CDM galaxy formation scenarios \citep{kauf}.
K-giants provide useful probes for understanding chemical evolution because
they are luminous, cover the full range of possible ages, contain lines
from many elements, and, except for a few light elements, they preserve the
composition of the gas from which they were formed.
The widely accepted paradigm \citep{t79,wst,t95,mcw97}
is that yields of massive star Type~II SNe are
enriched in alpha elements (e.g. O, Mg, Si, Ca, Ti), and the
longer-timescale Type~Ia SNe are enriched in iron.
Consequently, the relative importance of alpha elements and iron gives a
constraint on the enrichment timescale for the stellar population.
For example, \citet{e99} models bulge formation as a
maximum intensity star burst that concludes in $10^8$ yr.

The consequences for chemical evolution were modeled by
\citet{m99}, who predicted abundance trends in the bulge for a wide range of
elements.  The expectation is that the rapid formation
of the bulge should produce an over-enhancement of alpha
elements.  However, all elements are of interest because
different families of elements are thought to be made under a variety
of astrophysical circumstances, which can provide information on
the bulge environment during its formation, and may also lead to a greater
understanding of nucleogenesis.  Surveys of large samples of stars in 
globular clusters and dwarf spheroidal galaxies are just beginning, and our 
research will add the study of the bulge population to these efforts.

The earliest efforts to obtain spectra of bulge K-giants with digital
detectors were by \citet{wr83} and \citet{r88}.  These
studies concluded that there is a range of metallicity in the
bulge and found evidence for super-metal-rich stars.
R88 found the mean abundance for the bulge was [Fe/H]~$\sim~+0.2$.
\citet{mr94} obtained marginal high-resolution echelle 
spectroscopy (R$=17,000$, S/N$\sim$50) of 11 bulge K-giants from the 
sample of R88 using the CTIO Blanco 4-m telescope
and revised the abundance distribution in the bulge downward.  The results
of MR94 were used to re-calibrate the Rich (1988) low-resolution sample, 
and indicated a mean abundance of [Fe/H]=$-$0.25.  MR94 also found 
enhancements of the alpha elements
Mg and Ti, even for stars with the solar iron abundance.  The alpha elements
Si and Ca did not appear to be enhanced over the solar neighborhood trends; 
this was explained by MR94, in the context of the supernova nucleosynthesis 
predictions of \citet{ww95}, as due to a top-heavy bulge IMF.
These early studies were made difficult by the lack of accurate basic
optical and infrared photometry for the bulge, and the large, but uncertain,
reddening (now known to be highly variable).  The high metallicity and cool 
temperatures of
the bulge giants also posed difficulties due to blending, and because it was 
necessary to use lines with somewhat uncertain \gf-values.  In addition, the 
extreme stellar crowding in these low-latitude fields, and the faintness of 
the bulge stars, made it impossible to obtain the highest quality
data with 4m-class telescopes; as a result, the use of relatively low 
resolving power spectra led to considerable blending with CN lines.

The first spectrum
of a bulge giant with Keck \citep{c96} confirmed the metal-rich end
of the MR94 abundance distribution; however, the spectrum was of low S/N and it
was clear that a serious Keck telescope campaign was urgently needed.
This effort has been underway since August 1998.  Acquisition of
the spectra has proceeded slowly, since we require S/N$>50$ and
R$>45,000$; in fact, most of our spectra have R$=60,000$.

An analysis of a subset of the new Keck spectra was given in \citet{rm00}.
The analysis employed model atmosphere parameters based entirely
on the spectra rather than on photometry: effective temperatures were set by
demanding that Fe~I line abundances were independent of excitation potential,
whilst the stellar gravities were chosen so that Fe~I and Fe~II abundances were
equal.  The RM00 abundances at the metal rich end are $\sim$0.15~dex more metal
rich than MR94.  The key results of MR94 were confirmed for the alpha elements:
[Mg/Fe] and [Ti/Fe] are elevated, while Ca followed the disk abundance
trend.  Oxygen was clearly measured for the first time, and it was found
that [O/Fe] declines rapidly above [Fe/H]=$-$0.5, a result that we will 
confirm with new work presented in a future paper.

One problem with the RM00 study was due to the sensitivity of the spectroscopic
model atmosphere parameters to the adopted \gf-values of the iron lines.  In 
addition, the covariance between the spectroscopic temperature and spectroscopic
gravity meant that a slight error in temperature returned a large error 
in gravity, which then required an additional adjustment in temperature, in 
the same direction 
as the original error.  In this way it was possible to obtain spectroscopic 
parameters unexpectedly far from the true values. 

\citet{mr04} analyzed some of the same data as RM00 (and 
found here), but avoided the pitfalls found by RM00 by introducing two
spectroscopic methods for deriving \teff{} values.  The first method
derived temperatures based on the excitation of Fe~I lines, with \gf-values
obtained from the analysis of the well-studied K-giant Arcturus;
note that the Arcturus \gf-values include systematic effects, such
as the deficiencies of the adopted model atmosphere grid, and so are not
necessarily equal to laboratory values.
The second spectroscopic technique used for setting \teff{} in MR04 is
to force the ionization equilibrium between Fe~I and Fe~II lines using 
a gravity derived from photometric data and an assumed mass of 0.8~M$_{\odot}$. 
These two new spectroscopic techniques were relatively robust, as well
as insensitive to reddening uncertainties that might affect photometrically
determined temperatures.  The spectroscopic temperatures
from MR04 compared extremely well with \teff{} photometric temperatures
derived using the Alonso (V$-$K) calibration and (V$-$K) colors based on
2MASS K and OGLE V data.  In this work, we expand upon 
the improvements implemented in MR04.

The bulge stars studied here are located in Baade's Window, a region
of relatively low reddening (A$_{\rm V}~\sim~1.3$~mag) located about
4 degrees from the direction of the Galactic center.  If a galactocentric
distance of 8~kpc is assumed (e.g. Merrifield 1992, Carney et al. 1995),
then the closest approach of the line of sight
to the Galactic center is about 550~pc.  For comparison the scale height of
the bulge is approximately 350~pc \citep{wyse97}.  The star identifications we
use for the bulge stars are those of \citet{a65}.

In the present study we have employed two spectroscopic methods
for estimating stellar parameters, and we check these with V$-$K colors
derived from the newly available 2MASS K-band~magnitudes \citep{2mass}
together with V-band measurements from the OGLE microlensing survey 
\citep{sz96}.  These three methods
provide excellent agreement in the stellar parameters, so we are confident 
that the present study currently provides the best approach for the detailed 
abundance analysis of bulge giants.  We shall see that our results for the 
iron abundance scale now returns more closely to that of MR94 and MR04 
than RM00.

In Section~2, we describe the observations and data reduction.  In Section~3
we describe the basic problems in the analysis of metal-rich K-giants and
outline our solutions.  Section~4
contains a description of the new Fe line list needed for the analysis.
Our continuum measurement and equivalent width measurement method are
explained in Section~5, while the methods we use to obtain the stellar
parameters are described in Section~6.  The results of the analysis are
provided in Section~7, including a discussion of stars for which the
initial analysis failed.  Finally, in Section~8 we use our results to 
recalibrate low resolution observations of Baade's Window giants to rederive 
the metallicity distribution function.

\section{Observations and Data Reduction}

The observations of Baade's Window stars reported here were taken with 
the HIRES spectrograph \citep{vogt}
on the Keck I telescope between 1998 and 2001.  The spectra were taken
with the C1 (0.861 arcsec by 7.0 arcsec) and B2 (0.571 arcsec by 7.0 arcsec) 
slits yielding instrumental resolutions of 45000 and 60000, respectively.  
The higher resolution was used for the highest metallicity targets. 
The HIRES spectra roughly cover the wavelength range 5375 to 7875~\AA, 
with inter-order gaps increasing to the red.  The data were reduced with 
the MAKEE package from T. Barlow\footnote{MAKEE was developed by T. A.
Barlow specifically for reduction of Keck HIRES data. It is freely available on
the world wide web at http://spider.ipac.caltech.edu/staff/tab/makee/index.html
or http://www2.keck.hawaii.edu/inst/hires/makeewww/}.  
The reduced spectra have signal-to-noise (S/N) values between 45 and 100 per 
pixel.  The journal of Keck observations is given in Table~1, and sample 
sections of three stars are displayed in Figure~1.

Echelle spectra of nearby giant stars were obtained with the 
du Pont 2.5-m telescope at Las Campanas and the 0.6-m CAT telescope at Lick
Observatory.  These stars were selected from \citet{mcw90} to have similar 
M$_{\rm V}$, [Fe/H] and \teff{} values as the Baade's Window sample.
These data were taken to help check
the methods used for determining the stellar parameters of the bulge giants.
The du Pont echelle spectrograph has a resolution of about 30000, while
the CAT at Lick Observatory uses the Hamilton spectrograph \citep{v87} 
delivers a resolution of about 45000.  Both spectrographs
give complete wavelength coverage in the 5000-8000 \AA{} wavelength range.  
The HIRES
spectrum of $\mu$~Leo (HR~3905) was reduced using MAKEE, 
but the remainder of the stars
were reduced using IRAF\footnote{IRAF is distributed by
the National Optical Astronomy Observatories, which are operated by the
Association of Universities for Research in Astronomy, Inc., under cooperative
agreement with the National Science Foundation.} routines.  
The data from the Las Campanas 2.5-m
echelle were extracted, and scattered-light subtracted, using IRAF scripts
written by AM.  The S/N ratio of the disk spectra are very high, often over
200 per pixel.  Details of these spectra are included in Table~2.

The sample of disk giants includes a Lick/CAT observation of Arcturus.
In Sections~3 and~4 (and Tables~3 and 4) we use the \citet{hw00} Arcturus atlas
to help develop the analysis method.  In later sections, though, we analyze
the Lick/CAT spectrum of Arcturus (HR~5340 in the Tables~7 and~9) just like
the other disk stars.  In this way, the Lick/CAT spectrum of Arcturus provides
a consistency check showing how the lower resolution and S/N of the disk sample
affects the results as compared to the atlas.

The bulge sample was selected to cover a wide metallicity range in order
to investigate the chemical evolution of the bulge.  Baade's Window contains
stars from populations other than the bulge (namely foreground disk and 
background halo stars).  Fortunately, \citet{srt96} studied the relative
fraction of non-bulge contaminants within Baade's Window.  Figure~10 of SRT96
shows the probability a star being a member of the bulge, disk or halo as a
function of the observed V magnitude (V$_{\rm obs}$).  This probability
is a ``worst-case'' value for individual stars because it excludes any 
other information (color, metallicity, photometric parallax, etc.) that might 
be used to help further classify an individual star.

The bulge dominates the membership of stars with V$_{\rm obs}$ between
about 15.5 and 18.0, with the disk being the main contaminant at brighter
magnitudes.  Between V$_{\rm obs}$ of 16.5 and 17.5, the probability of 
bulge membership rises to over 80 percent due to the presence of the bulge red
giant branch clump in that magnitude range.  Background halo stars are
only a minor contaminant, with the maximum probability of contamination
rising to just over 10 percent around V$_{\rm obs}$ of 16.1 mag.  Halo
stars are the main contaminant in the regime of bulge RGB clump stars,
but it is likely the halo stars are confined to the lowest metallicities. 

For the stars observed here, the V$_{\rm obs}$ values place
only 10 stars in regions where probability of bulge membership is less 
than 80 percent.  Of those, all 10 are either individually identified 
by SRT96 as bulge members due to photometric parallax or were observed 
by MR94 and previously confirmed having distances consistent with bulge 
membership.  Therefore, in our initial analysis we can assume that all 
the stars we observed in Baade's Window are bulge members (we test this 
assumption in Section 7.1).  

It has been noted by SRT96 that velocity information is not very useful
as a population diagnostic in Baade's Window.  They find that the velocity
dispersions for the bulge, disk and halo in Baade's Window to be 103, 82 and 
149~km/s, respectively.  \citet{r90} found a bulge velocity dispersion of 
$104~\pm~20$~km/s.

\section{Difficulties in the Analysis of Metal-Rich Giants}

The bulge is distant ($\approx~8$~kpc) and heavily reddened 
(A$_{\rm V}~\sim~1.3$~mag in Baade's Window).  The only stars that
can be analyzed in large numbers at high-resolution and S/N are intrinsically
bright giants.  But the bulge is also metal-rich, with a mean metallicity
of roughly solar (see SRT96 and Section 7).  Bright, metal-rich giants suffer
from heavy line blanketing, and many of the well-studied lines with high 
quality laboratory \gf-values are very strong.  For example, the line list 
of \citet{f00} is based on Fe~I lines from the Oxford group \citep{b89}
and \citet{o91}.  When these lines are measured in the nearby metal-rich 
K-giant $\mu$~Leo, there are only 36 unblended Fe~I lines weaker than 
150~m\AA{} available in the wavelength region observed, and only 11 of 
those are weaker than 90~m\AA{}.

But there are also more basic difficulties affecting even the analysis of 
metal-poor giants.  For Arcturus, \citet{shm02}
found a disconcerting 110~K difference between the physical T$_{\rm eff}$ 
(based on the known flux, distance and angular diameter) and T$_{\rm eff}$ 
based on Fe~I excitation (\tex).  This difference was reduced to 
40~K if a group of high-excitation Fe~I lines was excluded from the
analysis, and suggests problems with the Fe~I \gf-values, or blends, at 
red wavelengths.  The remaining 40~K difference may be due to problems with 
the upper layers of the model atmospheres, where the empirical T-$\tau$ 
relation for Arcturus \citep{al75} differs significantly from 
the \citet{k93} model \citep{mcw95a}.

As an example of the problems encountered, we can conduct a simple analysis 
for the slightly metal-poor bulge star I-194.  First, we used the line
list of MR94 (with the addition of a few Fe~II lines from Fulbright 2000) and
measured 82 equivalent widths using the IRAF {\it splot} package and fitting the
continuum by eye.  For the purposes of this example only, we set the \teff{} 
value using the excitation method and set \logg{} by forcing ionization 
equilibrium between Fe~I and Fe~II.

When this is done, we find
\teff{}~=~4490 K, \logg~=~1.45, [m/H]~=~$-0.10$, and \vt~= 1.45~km/s,
which requires $M_{\rm Bol,*} = +0.95$, or a distance of 5.9~kpc--well
outside the bulge.  If we force a distance of 8~kpc, then the ionization
difference (\leps(Fe~II) $-$ \leps(Fe~I)) jumps to $-0.42$~dex.  
The photometric \teff{} (\tphot; see Sections 6.2-6.4 for details on how 
we determine stellar parameters) for this star is 4153~K, or 337~K cooler, and
if we assume a distance of 8~kpc, the ionization difference is +0.21~dex.  
Ionization equilibrium at this \teff{} requires a distance of
15.0~kpc, far on the other side of the bulge.  The range of [Fe/H] values
from all the above parameter determinations range from $-0.42$ to $-0.04$.  

Except for the wish to have the star located in the bulge, there are
no additional constraints on the parameters.  Both \tphot{} and \tex{}
methods are valid in theory, but the actual application here has led
to discordant results.  In Figure~2 we plot excitation plots from both
\teff{} methods based on the 8~kpc results.  In the lower panel, it appears
that the removal of certain lines might flatten the slope of the Fe~I
lines and bring Fe~II into agreement with Fe~I.  Indeed, the final \teff{}   
we adopt for I-194 is 4176~K, but there is no evidence here to support
the removal of those individual lines.  In other stars the lines that 
are outliers in this plot are near the mean, and other lines are outliers.
These lines also do not show any suspicious properties in the observed 
spectrum of I-194 that might explain their outlier status.

There are several reasons for the wide range of results for this star.
As mentioned in the Introduction, excitation temperatures and ionization-based 
surface gravities exhibit a covariance.  When \teff{} is raised, the 
difference \leps(Fe~II) $-$ \leps(Fe~I) decreases.  This forces an increase
in the surface gravity, which in turn changes the slope of the excitation
plot slightly, requiring another rise in \teff.

Further, the list has a large number of strong lines in this star.  There
are only 15 lines weaker than 50~m\AA{} and 20 stronger than 120~m\AA.
Abundances determined from strong lines depend greatly on the value of the
microturbulent velocity (\vt).  Of the 15 weaker lines, 13 have excitation
potentials of greater than 4~eV.  These high-excitation lines are sensitive
to \teff{} changes, so a change in \teff{} forces \vt changes.  One can
choose not to use stronger lines, but this greatly decreases the size of
the line list and makes the analysis vulnerable to a few pathological lines.
As mentioned in the example of $\mu$~Leo above, the number of weak lines
with high-quality \gf-values available rapidly decreases at high metallicity.  
Some of the bulge giants have metallicities over 3 times that of I-194.

Finally, the abundance results for individual lines display a sizable 
line-to-line scatter (about 0.20~dex for Fe~I and about 0.27~dex for Fe~II).
This kind of uncertainty can be due to errors in \gf-values, unresolved
blends in the spectrum (although MR94 attempted to use unblended lines),
and problems in continuum determination.  Uncertainties this large can 
reduce the diagnostic power of the parameter-determining methods, especially
if only a few lines are involved.

Problems like those detailed above convinced us to develop new techniques
to analyze the giants in our sample.  The need for more Fe lines and 
reduced line-to-line scatter drove us to create a new line list.  Worries
about the effects of blanketing on the continuum level induced us to 
create a new method to set the continuum.  And the inability to independently
select one \teff{} diagnostic as superior made us use three different
methods to set the parameters.

This development took several iterations to get to the final state,
which makes describing the methods in a linear fashion difficult.  
For example, we use stellar parameters (Section~6) to synthesize the
continuum regions (Section~5) so we can measure the Fe lines (Section~4),
but we need a good set of Fe lines before we can set the stellar parameters.  
In the end, we analyzed each star at least four different ways before 
performing a final analysis.

\section{Line List}

As stated above, one difficulty in the analysis
of very metal-rich giants is that many of the well-studied lines
with high quality laboratory \gf-values are very saturated, with only a mild
sensitivity to abundance.  Additionally, the weaker lines all have high 
excitation potentials, while the stronger lines are from lower excited states.
This sets up the possibility for correlated errors between the microturbulence
velocity (\vt) and the \teff{} value when the \teff{} value is determined by 
the slope of the \leps(Fe~I) versus excitation potential distribution (which we
will refer to hereafter as the ``excitation plot'').  This can be alleviated
by having a range of line strengths at all excitation potentials.

As mentioned in the previous section, McWilliam et al. (1995) also notes 
that Fe lines with strengths over 
log(RW) = log$_{10}({\rm W}_{\lambda}/\lambda)\sim -4.7$ 
($\sim 120$~m\AA{} near 6000~\AA) in very metal-poor stars
suffer from the effects of improper modeling of
the outer layers in many stellar atmosphere models.  We therefore
wish to exclude strong lines because we cannot rely on the abundance
results from these lines. 
In addition, strong lines are sensitive to the selection of the microturbulent
velocity.  We found that in the example of the \citet{f00} list and 
$\mu$~Leo, the mean abundance of the Fe~I lines could change by 0.1~dex for 
every 0.1~km/s change in \vt.  Due to the lack of weak lines, setting the 
value of \vt{} using the \leps(Fe~I) vs. log(RW) could lead to \vt{} 
uncertainties of up to 0.3~km/s in the stronger-lined stars.  This situation 
is unacceptable and a new line list with more weak lines is required.

Our goal was to create a list of Fe~I and Fe~II lines that span 
a range of line strengths and excitation potentials that are unblended
and are of appropriate strength ($\sim$~10--120~m\AA) for the range of
metallicities found in the bulge giants.  To find these lines, we
used the spectrum synthesis program MOOG \citep{moog} and the complete 
Kurucz line list ``gfall.dat''\footnote{The most recent versions of the 
Kurucz line list and atmosphere grids can be found at 
http://kurucz.harvard.edu/.  For the selection of the line list the 
solar-ratio grid with overshooting was used.  See section 6.1 for 
a comparison of how the choice of atmosphere model grid affects the final
results.}
combined with the CN line list from MR94 to calculate the estimated EW values 
in small ($\pm~2$~\AA) wavelength regions around over 5500 Fe~I and Fe~II lines 
between 5400--7900~\AA.  

We used stellar parameters to simulate Arcturus 
(\teff{}~=~4290 K, \logg~=~1.60, [m/H]~= $-0.50$, and \vt~=~1.67 km/s), 
and a hypothetical giant with parameters identical to Arcturus, but 
with [m/H]~=~+0.5.  

For the Arcturus \teff{} value we employed the limb-darkened
angular diameter of 20.91 $\pm$0.08~mas from \citet{p98},
which is in good agreement with the values in the range from
20.78$\pm$0.31 to 21.04$\pm$0.05~mas found by \citet{q96}.
To compute \teff{} from the angular diameter the bolometric flux is required;
we utilized four measurements of the bolometric flux of Arcturus,
from \citet{a80}, \citet{bg89}, \citet{a99}, and
\citet{b90}.  The median bolometric flux from these works is 
4.95$\times$10$^{-12}$~W/cm$^2$.  The Hipparcos catalog lists the distance 
to Arcturus
at $11.25~\pm~0.09$~pc; at this close distance it is safe to assume that there
is no significant interstellar reddening, although the existence of 
circumstellar reddening is difficult to rule out.  The median bolometric flux, 
listed above, indicates \teff = 4294~K for Arcturus.  The \citet{a00} 
bolometric flux value is significantly lower than all other estimates, and 
indicates 
\teff = 4268~K, while the highest bolometric flux is from \citet{bg89}, 
consistent with \teff = 4337~K.  Thus, we adopt a \teff{} value of 4290~K, 
close to the value of 4280~K determined by \citet{mcw90}; we estimate the 
1$\sigma$ at $\sim$10 to 15~K.

The absolute luminosity of Arcturus was determined from the observed V-band 
magnitude, bolometric corrections, and the Hipparcos distance of 11.25~pc.  
The mass was estimated at 0.90 \msun{} from the location of Arcturus in the 
temperature-luminosity diagram, by comparison with the Padova theoretical 
isochrones \citep{gir00}.  From Equation~1 and the isochrone mass we compute 
the surface gravity for Arcturus, at \logg = 1.55, rounded to 1.6.

Our choice of parameters for Arcturus agree with other analyses 
\citep{glg99,p93}.

In our syntheses,
Kurucz model atmospheres (interpolated from the grid points using a program
given to us by J. Johnson) with solar abundance ratios were used, but the
abundance of [C/Fe] and [N/Fe] were changed to be $-0.2$ and +0.4~dex,
respectively; this is typical of solar neighborhood red giant stars, which
have altered their surface composition via dredge-up of material that has
undergone nuclear processing in the stellar interior 
(e.g. Lambert \& Ries 1981).
The change in the [C/Fe] and, especially, [N/Fe] ratios affects the strength
of CN features in our spectra.

The vast majority of the Fe lines considered are too weak, too strong,
or too blended with other lines to be useful in our abundance analysis.
Selected lines must have a simulated EW of over 2~m\AA{} in 
Arcturus and less than 150~m\AA{} in the [Fe/H] = +0.5 giant.  We consider   
a line too blended if the sum of the lines within 0.2~\AA{} of the center of
the Fe line 
contribute more than 0.05~dex to the resulting log(RW) value of the blend.
Further, we inspect the actual spectrum of $\mu$~Leo to ensure that no 
unaccounted lines in the synthesis contaminate the Fe line of interest.  
We note that our list of clean Fe~I lines is restricted by the coverage of 
the $\mu$~Leo spectrum; additional lines may be identified in later studies.  
The line list presented here is made up of the Fe lines with the least amount 
of stellar or telluric contamination possible in the wavelength regions 
observed.  The final result is a list of 154~Fe~I lines and 5~Fe~II lines.  

Most of these lines, however, do not have laboratory \gf-values available.
We have therefore chosen to conduct our analysis differentially from 
the well-studied giant Arcturus.  A differential analysis also helps reduce 
systematic errors, such as deficiencies in the model atmospheres
and the effects of weak blends that scale with metallicity.

The differential abundance analysis was performed on a line-by-line basis for 
a subset of our Fe lines, which we believe to be least affected by blends,
with strengths under 120m\AA{} in the Sun and Arcturus.  
The EW values of lines in the Sun and Arcturus were measured from the 
high-quality spectra of \citet{ksun} and \citet{hw00}, respectively, 
and are listed in Table~3.  Table~3 contains a subjective quality assessment
for each line, with A the highest quality and E the lowest.

In order to perform the differential abundance analysis it is necessary to 
adopt a microturbulent velocity value for the atmospheres of both stars.  A
sufficient approximation would be to choose a microturbulent velocity value
for the Sun from the literature. 
However, we chose to estimate the microturbulent velocities for the Sun
using an absolute abundance analysis, from a small sample of Fe~I lines with 
accurate laboratory $gf$-values.  
We could then select the Arcturus microturbulent 
velocity by demanding that the differential abundances are independent of EW.
Our adopted $\log{gf}$ values are unweighted 
averages from the work of three groups: \citet{o91}; \citet{b91} and 
\citet{b94}; and \citet{ox4,ox5,ox1,ox2,ox3}.
All three groups claim small uncertainties, near 0.04~dex, for most of their 
lines.  Small systematic differences in the $log~gf$ values from the three 
sources indicate the values of the O'Brian group are systematically larger 
than the Bard group and both groups have values larger than the Blackwell 
group; however the mean difference between the O'Brian and Blackwell groups, 
at only 0.03~dex, is comparable to the random uncertainties, and we have, 
therefore, chosen not to attempt systematic corrections of the $log~gf$ scales.
The adopted $log~gf$ values and EWs in the Sun of these Fe~I lines 
used for the absolute abundance analysis are listed in Table~4. 

In the calculation of the abundances and microturbulent velocities we adopted 
solar atmosphere parameters of T$_{\rm eff}$=5770~K, logg=4.44, [m/H]=0.0, and 
Arcturus parameters of T$_{\rm eff}$=4290~K, logg=1.55, [m/H]=$-0.50$.  The most
appropriate models are the models including the new opacity distribution 
functions from Fiorella
Castelli\footnote{http://wwwuser.oat.ts.astro.it/castelli/.},
ODFNEW, for the solar model, and the alpha-element enhanced models 
with the new opacity distribution functions, AODFNEW, for Arcturus.  
With these models, and the best laboratory Fe~I $gf$ values listed in Table~4,
we found microturbulent velocities for the Sun and Arcturus of
0.93 and 1.67 km/s, respectively.  The absolute iron abundances, computed using
the Fe~I lines from Table~4 are 7.45~dex and 6.95~dex for the Sun and Arcturus
respectively.  Note that since the iron abundances of our stars are computed relative
to Arcturus, they are not sensitive the absolute solar iron abundance.

We calculated the line-by-line differential iron abundance of Arcturus, 
[Fe/H], relative to the Sun for a variety of model atmosphere grids.  In the optimal
case presented above, the line-by-line difference is $-$0.50 ($\sigma$=0.07).
If ODFNEW models are used for both stars the difference is $-0.56~\pm~0.07$, while
the difference is $-$0.51 $\pm$ 0.07 when AODFNEW models are used for both stars.

It is impressive that the [Fe/H] derived
from Fe~I lines is fairly insensitive to the choice of model atmosphere grid. 
In contrast, the iron abundance derived from Fe~II lines is quite
sensitive to the model atmosphere choice (for the three cases presented above
the differences are $-0.44~\pm~0.04$, $-0.61~\pm~0.04$, $-0.54~\pm~0.04$,
respectively) in a manner expected from the sensitivity of lines
from ionized species to the electron pressure.

Although it would be possible to estimate $log~gf$ values for our clean Fe 
lines in Table~3, we choose not to do so because of the strong dependence 
on the input assumptions, such as the model atmosphere grid used, the method
for interpolation within the grid, and the spectrum synthesis program used for
the analysis.  Readers who wish to use this 
list of relatively unblended Fe-lines are advised to either perform a
differential analysis themselves or to create astrophysical \gf-values.

\section{Continuum and Line Measurements}

Line measurements in very metal-rich giants require great care because
of the difficulty of determining the continuum level in highly blanketed
regions.  For our highest resolution spectra of bulge giants, a 1\% 
change in the continuum placement will yield a systematic 4.5~m\AA{} 
change in the equivalent width (EW) values.  The effect of an additive shift
to individual lines depends on the original line strength.  A 4.5~m\AA{} 
increase of a hypothetical 120~m\AA{} line at 7000~\AA{} increases the 
resulting abundance by about 0.02~dex, but the effect of the shift on the 
abundance determined from a 10~m\AA{} line is about 0.16~dex.  The increased
effect on weaker lines will affect the determination of the microturbulent
velocity and excitation temperatures.  Therefore we needed to employ
a method to accurately determine the continuum level.

The spectra extracted by MAKEE, and IRAF scripts, were first roughly flattened
by dividing by a blaze function spectrum.  The blaze function was determined 
from fitting the continuum of a star with very few lines, typically a very 
metal-poor star, using the IRAF {\it continuum} script.  Unfortunately, the 
blaze functions from the HIRES spectra were not very stable, and may have 
depended upon where in the sky the telescope was pointing; however, since 
several blaze function candidates were available, we chose the blaze function 
which gave the flattest normalized object spectrum.  We note that the IRAF 
{\it continuum} fitting parameter variables are somewhat subjective in this 
method.  For example, weak line blanketing, of order a few percent, if not 
accurately removed can introduce subtle ripples into the blaze function, that 
can create artificial high-points in the normalized spectrum.

Continuum fitting was performed using the semi-automated spectrum measuring
program GETJOB \citep{mcw95b}.  Although GETJOB has options to automatically 
select
continuum regions based on high points in the spectrum, we chose to use the
mode where the continuum region wavelength bounds are given as inputs
to the program.  GETJOB would then fit the continuum fluxes, as a function of
wavelength, using a polynomial least-squares fit.  The order of the fit was
automatically selected to minimize the chi-squared residual normalized by the 
number of degrees of freedom; typical fits were order 3 or 4 polynomials.

The input continuum regions were selected on the basis of spectrum synthesis
experiments, including a spectrum synthesis of the very metal-rich red 
giant $\mu$~Leo, over the 5000--8000 \AA{} wavelength range.  We used the 
same Kurucz line list, with added 
CN lines, used in the creation of the Fe line list discussed in section~4. 
The synthesized spectrum was then searched
for regions where the normalized flux stayed above 0.99 for more 
than 0.3 \AA{} (about 8 pixels in the HIRES spectra).  These regions were 
then visually inspected in the observed $\mu$~Leo spectrum to confirm that 
that synthesis accurately reflected the local stellar spectrum and to 
ensure that no telluric or unidentified stellar lines were within
that region.  In the bluer regions, where the line blanketing is larger, 
we were forced to lower the threshold to 0.95, in order to have sufficient
numbers of continuum regions per order.  The list of continuum regions employed 
in this work are presented in Table~5.  Table~5 also gives predicted relative
flux values for
these continuum regions in five stars:  the Sun, Arcturus, $\mu$ Leo, and the
bulge giants I-025 and IV-003.  These last two stars are respectively the most 
metal-rich and metal-poor stars in the bulge sample.

Prior to measuring line EWs in each star, we synthesized the flux for every 
continuum region using the Kurucz line list plus our CN list and an atmosphere
with estimated stellar parameters (based on the photometric \teff, see Section 6.1).
The mean relative flux value within each continuum region was calculated
and used as a correction factor within the GETJOB 
line measuring program.  The GETJOB continuum fit would include a 
re-normalization for each continuum region, based on the predicted continuum 
line blanketing factor,
in order to estimate the true continuum level.
For the very metal-rich bulge giant I-025, the mean value of the 208
continuum regions is 0.994.  During the interactive fitting
process, the observed spectrum within each continuum region was visually 
compared to the synthesis to ensure that no telluric 
features had been velocity-shifted into the region.

With the continuum level set GETJOB automatically measured the EWs of the 
spectral lines, from least-squares Gaussian fits to the line profiles.  
GETJOB includes routines that search for line blends, based on the S/N of the 
spectrum.  GETJOB also allows for fully interactive fits, where the user sets 
the pixel range over which
the Gaussian fits are made; this interactive facility was used to check the EWs
of all lines manually.  The EW values for all lines, in all stars, of our 
bulge and solar neighborhood samples are presented in Table~6.

In Figure~3 we present an example of what can go wrong when mis-using an
automated continuum algorithm that employs spectrum high points to set the 
continuum level.  If the number of pixels summed is too small the continuum of
metal-poor stars is set by statistical high points and peaks in the ripple of 
the blaze function, and other systematic errors.  On the other hand, for very 
metal-rich stars the continuum regions are short and rare, so that the number 
of pixels summed is too large, and thus includes regions affected by line 
blanketing, resulting in
underestimated continuum levels.  Although these effects may result in continuum
levels with typical errors of only a couple of percent,
this size of error is enough to affect the derived line abundances, especially
for weak lines, which are crucial to the determination of the microturbulent
velocity.  

Figure~3 shows that for photometric and ionization temperatures the systematic 
abundance errors, induced by continuum setting problems, ranging from 
$+0.05$~dex from metal-poor stars to $-0.05$~dex for very metal-rich stars.  
The abundance differences are greater if the temperature is set by Fe~I 
excitation than for the ionization and photometric 
\teff{} methods.  This may be due to a correlation between line strength and 
the excitation potential of the line: weak lines, generally of higher 
excitation, are affected more by an additive offset from a continuum change.  
These lines will then be systematically lower or higher (depending on the
direction of the continuum shift), affecting the slope of points on the
excitation plot.  Indeed, for the 11 stars in Figure~3 with [Fe/H] $< -0.5$, 
the excitation temperatures for the synthesis method are on average  
$121~\pm~48$~K (standard deviation of the mean) cooler.  This is enough 
to explain the lower abundances derived for the synthesis method measurements.
Since ionization temperatures depend on the mean of many Fe lines rather than
a slope, the effect is less sensitive to the weak lines, and so greatly 
diminished.  Note that the change in abundances for ionization temperatures 
follows the same change seen in photometric temperatures, which means the 
changes in [Fe/H] seen is due to equivalent width and \vt{} changes.

\section{Atmospheric Parameters}

The high and variable extinction toward Baade's Window \citep{s96,f99},
and the possibility of variation in the ratio of total to 
selective extinction \citep{u03}, increases the difficulty of determining 
stellar parameters of our bulge stars by purely photometric means.  This is 
complicated by the lack of accurate distances for individual bulge stars.  
Given these uncertainties, and the potential for error in the abundances, we
elect to calculate the stellar atmospheric temperatures using three 
independent methods:  photometric color-temperature relations; temperatures 
based on the excitation of Fe~I lines, with abundances taken relative to 
Arcturus; and temperatures based on ionization equilibrium of iron, again with 
abundances taken relative to Arcturus, and a gravity adopted from 
photometric data.

\subsection{Choice of Atmosphere Model Grid}

The use of an appropriate stellar atmosphere grid is important to
high-quality abundance analysis.  For example, in these stars, the continuous 
opacity is dominated by H$^{-}$.  The formation of H$^{-}$ depends greatly on 
the supply of free electrons contributed by metal atoms, especially Mg.  
Previous analyses (MR94, RM00, MR04, etc.) have found that bulge stars of
all metallicities have enhanced Mg/Fe ratios as compared to the Sun, but 
not all of the other alpha-element ratios are similarly enhanced.  
Therefore, it is important to understand the effect of using either 
solar-ratio or alpha-enhanced model grids.

Similarly, there may be other assumptions (such as whether to use grids
with overshooting on or off) that affect the final results.  The primary
analysis in this paper was done using Kurucz model 
grids\footnote{http://kurucz.harvard.edu/} employing
solar abundance ratios with overshooting turned on.  

To determine how the assumptions of the different atmosphere grid types affects
our abundance results we have repeated a subset of the abundance analyses with
different atmosphere grids.
In the investigation the line measurements and parameter-setting methods for the
Arcturus and bulge star abundances remained unchanged; however, we note that the
Arcturus analysis used in the line-by-line differential analysis is based on
the same grid of atmospheres.  We know that Arcturus is alpha-enhanced (Peterson et al. 1993;
Fulbright, Rich \& McWilliam 2005 in preparation),
so the most appropriate model for an absolute abundance analysis is an 
alpha-enhanced model atmosphere.  We note that our differential analysis, relative
to Arcturus, roughly compensates for deficiencies in the models.  This 
correction should be most accurate for stars having compositions and physical parameters
similar to that of Arcturus.

The atmospheres included in the tests include Kurucz models with overshooting
turned off (``NOVER'' taken from the same Kurucz web site as the overshooting
models), grids from Fiorella 
Castelli\footnote{http://wwwuser.oat.ts.astro.it/castelli/.}
using new opacity distribution
functions assuming solar abundance ratios (``ODFNEW'') or alpha-enhanced
abundance ratios (``AODFNEW'').  Finally, we included ``MARCS'' models 
\citep{marcs} calculated by a binary executable file provided by M. Shetrone.  

We present the results from the tests in Table~7.  Each column of numbers 
gives the mean value of the difference of that parameter for the model
grid in question minus the result from solar-ratio overshoot grid (our
default grid).  We calculated the mean difference for the entire sample
and for a subsample of stars with [Fe/H] $> 0$.  We tested the effect
on the derived excitation and ionization \teff{} values (see Section 6.4), 
the derived Fe~I abundance for all three \teff{} scales, and the ionization 
difference (\leps(Fe~II) $-$ \leps(Fe~I)) for the photometric
and excitation \teff{} scales (the ionization \teff{} scale is defined
by forcing the ionization difference to zero).  The value of $\sigma$ 
given in the table is the standard deviation of the results.  

The difference caused by the use of alpha-enhanced models was small:
When photometric \teff{} values were used, there was little
change in the derived abundance from Fe~I lines, but the ionization 
difference decreased by 0.03~dex--the abundance derived from Fe~II lines
is somewhat sensitive to increasing the alpha-abundance of the atmosphere.
The effect is larger in the more metal-rich stars ($-0.07$~dex).  It is
not clear whether these alpha-enhanced models (where all of the alpha element
abundances are increased by $+0.4$~dex) are any more representative of 
the metal-rich bulge stars than the solar models.  Previous analyses
of metal-rich bulge stars have found enhanced Mg ratios, 
important to the supply of free electrons, and solar-ratio abundances
of Si and Ca, which are important to the mean molecular weight of the
atmosphere.  It will probably be necessary to re-address this issue
during the analysis of the alpha elements in a future paper.

As a warning, these tests do not show that solar-ratio or
alpha-enhanced model grids are roughly interchangeable when
applied to absolute abundance analyses, but only that a 
differential analysis can somewhat compensate for the use
of atmosphere grids calculated with different input assumptions.

\subsection{Photometric Data}

All three methods used to derive the stellar model atmosphere parameters require
some photometric inputs.  
One of the great improvements in observational data since the MR94
study is the availability of 2MASS JHK \citep{2mass} and OGLE VI 
\citep{sz96} photometry for the bulge.  \citet{ws96} and \citet{s96} have
also used the OGLE photometry to create a position-dependent reddening map
for Baade's Window, allowing for reddening estimates for each 
of our stars.  The available OGLE data includes data for 25 of our 27 stars.  
For the two other stars (IV-047 and IV-167) TSR95 V~magnitudes were used.
Of our sample, only IV-047 does not have data in the 2MASS Point Source 
Catalog.  It was necessary to transform the 2MASS data from the 2MASS filter 
system to the TCS filter system used by the \citet{a99} Teff-color relations
using the transformations available on the 2MASS web site\footnote{http://www.astro.caltech.edu/~jmc/2mass/v3/transformations/} and \citet{a98}.  Similarly,
the reddening measures of \citet{s96} were based on the SAAO filter system, 
so the E(B$-$V) measures were first transformed into the 2MASS system and 
removed from the data before any other transformations were applied.  
We list the final dereddened photometric data in Tables~1 and 2.  We
assume $A_{\rm V}~=~3.136~{\rm E(B-V)}$ from the reddening law of \citet{w97}.  
The final \tphot{} values 
were calculated using Equation 9 of \citet{a99}.  

For the comparison disk sample, the V and B$-$V photometric data come from 
the Bright Star Catalog \citep{bsc}, while the V$-$K data is from \citet{jp11} 
as found on 
SIMBAD.  These stars are too bright for 2MASS to provide reliable
photometry ($\sigma_{\rm K} \sim$ 0.3~mag), so we were 
forced to adopt the T(B$-$V) when V$-$K 
data were not available.  Since all stars in the present disk 
sample were also studied by \citet{mcw90}, we can use the \citet{mcw90} 
extinction estimates to correct the photometry: only HR~2113 was found to have
any evidence for extinction, with A$_{\rm V}$=0.17 magnitudes.  
The \tphot values derived for our 17~disk stars are on 
average $50~\pm~29$~K cooler that the values derived in \citet{mcw90}.
The final photometric \teff{} values are given in Table~7.

\subsection{Distances and Surface Gravities}

We can determine the surface gravity of a star if we know its mass ($m$), 
temperature, and bolometric~magnitude ($M_{\rm bol,*}$) using 
the following well-known equation:

\begin{equation}
\log{g} = \log{(m/m_{\odot})} - 0.4(M_{\rm Bol,\odot} - M_{\rm Bol,*}) + 4\log{(T/T_{\odot}}) + \log{g_{\odot}}
\end{equation}

We adopt $M_{\rm Bol,\odot}~=~4.72$, $T_{\odot}~=~5770~K$ and 
$\log{g_{\odot}}~=~4.44$.  In the above equation, 
the uncertainty in \logg{} from distance errors scales roughly as 0.4 times 
the uncertainty in $M_{Bol,*}$ (which folds in errors in the bolometric 
correction, distance modulus, photometry, and reddening).  

As stated before, there are no definitive distance determinations
for any of our Baade's Window sample stars.  As discussed
in Section 2, previous work shows that the individual probability that each
star in our sample is a bulge member is very high.
We can show that we do not need very accurate distances in order to determine 
good surface gravity values.  We adopt a distance to the Galactic Center of
8.0~kpc, based on the work of \citet{eisen} and \citet{r93}, which
corresponds to a distance modulus of 14.51~mag.  A $\pm~0.5$~mag shift in 
the distance modulus changes the distance to either 6.3 or 10.0~kpc.  In other 
words, a small distance modulus error quickly locates the star outside the 
bulge.  A 0.5~mag error in $M_{Bol,*}$ leads to a 0.12~dex error in \logg, 
which has only a small effect (about 0.05~dex) on the final ionization 
equilibrium.

Therefore, we can assume with some certainty that all the bulge stars in our 
sample lie at a distance of $8.0_{-3.7}^{+2.0}$~kpc.
If the resulting abundances point to surface gravity selection problems 
(e.g., a large disagreement in the abundances derived from Fe~I and Fe~II), 
then it is very likely the star in question is not a bulge star.

For the disk sample only, we used the Hipparcos parallaxes \citep{hip} to
determine the distance to the stars.  For most of these stars the parallax is
known to better than 10\%.  The distances to the disk giants range from
about 11~pc for Arcturus to about 145~pc for HR~1585.

\subsection{Parameter Determination}

The abundance analysis was conducted by a suite of programs customized
to work with the MOOG equivalent width abundance analysis program.  These 
programs applied the line-by-line differential analysis with respect to 
Arcturus and through an automatic iterative process would determine the 
best model atmosphere for the chosen method.  

The main difference in the three parameter-setting methods is how
the \teff{} value is determined.  The other major parameters (\logg,
[m/H], and \vt) are set in the same way in all three methods.  Before
we discuss the differences in the temperature-setting methods, we will
discuss how the other three parameters are set.

As mentioned above, we determine the \logg{} value for all cases using 
Equation~1.  For the Baade's Window sample, we adopt a
distance of 8~kpc and a stellar mass of 0.8 M$_{\odot}$.  For the disk sample,
we use the Hipparcos distance and derive a mass by the location of the
star in the $M_{\rm Bol}$ vs. log \teff{} diagram when compared to
Padova evolutionary tracks \citep{gir00} of a similar metallicity.  
We assume M$_{\rm Bol,\odot}$ = 4.72, T$_{\odot}$ = 5770 K, and 
$\log{g_{\odot}} = 4.44$.

The bolometric corrections, BC(V), for our stars come from \citet{a99}.
Our analysis program automatically recalculates new BC(V) and \logg{} 
values whenever either of the input variables (\teff{} or [Fe/H]) changes. 
If we adopt BC(K) instead of BC(V) the final M$_{\rm Bol}$ (see Section~7.2 and 
Table~8) is increased by $+0.04 \pm 0.08$ mag; presumably this is the scale
of the uncertainty of the bolometric corrections.  Because the V-band suffers
from greater line blanketing than the K-band in cool stars, we assume 
that BC(V) values are more sensitive to line formation; thus, we expect 
greater uncertainty in BC(V) for cool stars and stars with high metallicity 
or unusual composition. The difference between M(K)$_{\rm Bol}$ and 
M(V)$_{\rm Bol}$ for II-122 is -0.27 mag.  This star is closest to the
RGB tip than any star in the Baade's Window Sample.  Use of
M(K)$_{\rm Bol}$ instead of M(V)$_{\rm Bol}$ changes the final
\logg{} value by only $-0.11$~dex.
There are only two other cases where the difference in bolometric magnitude
exceeds 0.10 mag: two bright Baade's Window giants (I-322 and IV-203), 
both with M$_{\rm Bol} < -2.25$, show differences of $+0.12$ and $+0.11$ mag, 
respectively, between the M$_{\rm Bol}$ values derived from the K- and V-band.  
Therefore, it is likely the bolometric corrections
become more uncertain with increasing luminosity, but the size of the 
uncertainty is not enough to cause major changes to the final abundances.

The microturbulence value (\vt) was set by the slope of \leps(Fe~I) vs.
log(RW) plot.  The final \vt{} value was set by forcing the slope of a
least-squares fit to the plot to be zero.  The uncertainty in the final
value can be estimated by the formula:
\begin{equation}
  \sigma({\rm v_t}) = \sigma(\log\epsilon{\rm (FeI)}) \left(\frac{\delta (\log\epsilon{\rm (Fe~I))}}{\delta({\rm v_t})} \right)^{-1}
\end{equation}
where the $\sigma(\log\epsilon{\rm (FeI)})$ value is determined from the standard
deviation of abundances of individual Fe~I lines and 
$\delta (\log\epsilon{\rm (Fe~I))}/\delta({\rm v_t})$ is determined empirically by the 
analysis program.
The atmospheric model [m/H] value was chosen to match the 
$\log\epsilon{\rm (Fe~I)}$ value determined by the previous iteration.

The ``photometric temperature'' (\tphot) method is the simplest.  The final
\teff{} value is the one determined by the \citet{a99} 
(V$-$K)-\teff{} relationship.  The main problem with this method is
the possibility of uncorrected differential reddening.  An error
of 0.1~mag in E(B$-$V) will yield a change in \teff{} of about 100 K
in a 4000 K giant.  The results from the \tphot{} determinations were
used as the initial conditions in the iterative method used to determine
the excitation and ionization temperatures.

The availability of 2MASS colors means we could also derive photometric \teff{}
values from a (J$-$K)-\teff{} relationship.  The use of two infrared bands  
reduces the sensitivity to reddening, but unfortunately the J$-$K color is not
as sensitive to temperature changes as V$-$K.  The \citet{a99} (J$-$K)-\teff{} 
relationship has an internal calibration uncertainty about five times larger
than the (V$-$K)-\teff{} relationship (125K versus 25K).  When applied to 
the Baade's Window sample, the mean difference of T(V$-$K) $-$ T(J$-$K) is
$-19 \pm 113$ K (standard deviation).   The quoted standard deviation of the 
difference is consistent with the uncertainty of T(J$-$K) calibration.

The ``excitation temperature'' (\tex) method requires that we adjust the 
\teff{} value of the star until the slope of the \leps(Fe~I) vs. excitation 
potential (EP) plot is zero.  This method (and the ``ionization temperature'' 
method below) reduces the potential problems due to reddening errors, but
is not free from uncertainty.  As mentioned earlier, in many of the
stars the strong lines are also those with low EP and the weak
lines are those with high EP.  Our new line list reduces this
problem somewhat, but in many stars the number of low excitation
(EP $< 3$ eV) lines was small.  Any problems with these lines could
systematically cause errors in the \teff{} values.  We estimate
the \teff{} uncertainty by:
\begin{equation}
  \sigma(\log{\rm T_{Ex}}) = \sigma(\log\epsilon{\rm (Fe~I)}) \left(\frac{\delta (\log\epsilon{\rm (Fe~I))}}{\delta({\rm T_{eff}})} \right)^{-1}
\end{equation}

Finally, the ``ionization temperature'' (\tion) method requires ionization
equilibrium between Fe~I and Fe~II.  Normally in spectroscopic parameter
determinations the value of \logg{} is altered to force ionization equilibrium.
In this method, the value of \logg{} is forced by the distance and mass
(with a slight dependence on \teff) by Equation~1, so the main way to 
affect the ionization equilibrium
is to change the \teff{} value.  A $+100$~K change in \teff{} changes
the value of $\log\epsilon{\rm (Fe~II)} - \log\epsilon{\rm (Fe~I)}$ by 
about $-0.1$~dex.  

The ionization temperatures method is insensitive to the effects of a few 
pathological Fe~I lines, 
but the scarcity of good Fe~II lines is a concern, as any errors  
in the measurement and
analysis of those lines could greatly affect the result.  The sensitivity
of the ionization equilibrium to \logg{} also means that distance errors
could lead to large systematic errors in \teff{} by this method.  We estimate
the internal \teff{} uncertainty by:
\begin{equation}
  \sigma({\rm T_{Ion}}) = \sigma(\Delta\log\epsilon{\rm (Fe)}) \left(\frac{\delta (\Delta\log\epsilon{\rm (Fe))}}{\delta({\rm T_{eff}})} \right)^{-1}
\end{equation}
where
\begin{equation}
  \Delta\log\epsilon{\rm (Fe)} = \log\epsilon{\rm (Fe~II)} - \log\epsilon{\rm (Fe~I)}.
\end{equation}

\section{Results}

\subsection{Results of the Individual Methods}

As described in the previous section, each of the three methods for determining
the stellar temperatures 
have their own strengths and weaknesses.  All three methods can give good
results when the input data is accurate (as the results for the disk sample
show below), so we have no initial preference for any of the methods.
Our use of three methods for estimating \teff{} ensures that our
temperatures are robustly determined, and allows for us to check for
systematic errors for individual measurements.

There are a few steps common to all three methods.  Errors in those steps
could lead all three methods to return incorrect results.  For example,
the effect of using different stellar atmosphere grids was discussed in 
Section~6.1.  Our surface gravities rely on several inputs:  \teff, 
mass, and bolometric magnitude.  The bolometric magnitude depends on 
the distance, photometric magnitude, and bolometric
correction.  Distance errors will have a large effect on the ionization
\teff{} values, and we discussed the effect of filter choice on the 
bolometric correction in Section~6.4.  

If all three methods are valid and all the input data and assumptions
are correct, the three \teff{} values obtained should all agree and
the stars should be in ionization and excitation equilibrium.  In Figures~4 
and~5, we plot the differences in \teff{} determinations and between Fe~I and
Fe~II abundances, respectively.  

In both plots the disk stars show excellent agreement between all three
\teff{} methods.
This shows that the methodology is sound when we know the
star's distance and reddening very well and the S/N level is very high.
In addition, the disk star abundances agree with previous determinations.
The mean [Fe~I/H] for the 17 disk stars from this work are 
$+0.07~\pm~0.02$~dex greater than found by \citet{mcw90} (for the 11 stars below solar
metallicity the mean difference is $+0.03~\pm~0.02$~dex).
The giants HR~1346 ($\gamma$~Tau), 
HR~1409 ($\epsilon$~Tau) and HR~1411 ($\theta$~Tau) are all members of
the Hyades cluster (vB~28, vB~70, and vB~71, respectively).  The mean [Fe/H] 
value for the three stars is $0.17~\pm~0.02$.  
\citet{bf} and \citet{p03} studied solar-type Hyades dwarfs (which avoids 
many of the analysis problems encountered here) and found a mean [Fe/H]
values of $+0.127 \pm 0.022$ and $0.13~\pm~0.01$, respectively.  
Given typical systematic abundance analysis scale uncertainties, of approximately 
0.05~dex, we conclude that the abundances determined here are consistent with 
previous studies.

For the most part, the temperature values found for the bulge stars 
show excellent internal agreement as well.
There are, however, six stars that do not show agreement in all of
the various comparisons.  The metallicity of these stars does not show
any bias to high or low [Fe/H] values.
Fortunately, we can diagnose the potential failures of the methods.
For example, two of the six stars (IV-047 and II-172) show \tphot{}
values that are much lower than the \tex{} and \tion{} values,
even though the later two are in agreement.  The most likely source of
the problem is poor photometry in the crowded field of Baade's Window.
For IV-047 there was no entry in the 2MASS Point Source Catalog, nor
any other published JHK colors.  Therefore, we were forced to use the 
photographic B$-$V colors of \citet{a65} and the T(B$-$V) calibration of
\citet{a99}.  The typical uncertainty in the \citet{a65} B$-$V colors of
$\sigma \sim 0.14$ magnitudes near V~$\sim~17$ (e.g. see van den Bergh 1971)
corresponds to 1-$\sigma$ \teff{} uncertainty of $\sim200$~K; thus
we give low weight to the photometric \teff{} for IV-047.

The sample of SRT96 includes both II-172 and IV-047.  They give V$-$I colors
for both stars on the Cousins system.  We use the color transformations of 
\citet{b79} and E(V$-$I)$_{\rm C}$~=~1.26~E(B$-$V)
to convert the colors onto the Johnson system for use in the
color-\teff{} relations of \citet{a99}.  This yields \teff{} values of
4393~K for II-172 and 4433~K for IV-047.  Both of these values are much
closer to the \tex{} and \tion{} values than the original \tphot{} values.

For II-172, 2MASS data are available, but it is possible that crowding
affected this star due to the relatively large pixels used in that survey.  
On the finding chart of \citet{a65}, II-172 is located very close 
to a much brighter star; II-172 is the only star in our sample that suffers 
from this problem.  The \citet{a99} T(J$-$K) for II-172 is 4281~K, which  
is close to the T(V$-$K) value of 4260~K.  However, the T(J$-$K) relation of 
\citet{a99} has an internal uncertainty five times larger than the
T(V$-$K) relation used here.  
Differential reddening is unlikely because the large
\teff{} difference ($\sim200$ K) would require a change of E(B$-$V) of over
0.2~magnitudes.  

A change in the distance modulus to either star would not solve the
discrepancy.  A 1 magnitude change in the distance modulus leads to a
$+65$~K change in \tex{} and a $-150$~K change in \tion.  A distance
change for these two stars could bring either \tion{} or \tex{} in agreement 
with \tphot, but not both.  Below we will use a distance change to correct
cases where \tion{} is much different than \tex{} and \tphot.
Therefore, since we cannot find reliable \tphot{} values for 
either IV-047 and II-172, we will ignore the \tphot{} results for these
two stars.

Another of the six ``problem'' stars is I-322.  In this case, the \tex{} 
value is out of line with the other two methods.  Under close inspection,
the excitation plot of this metal-poor star contains only one line with
an excitation potential of less than 1 eV.  The least-squares fit is 
very sensitive to outliers, so it is possible that the real uncertainty 
of the \tex{} value is larger than the formal value of
37 K.  No other simple solution would solve the problem, so we will ignore
the \tex{} results for this star.

The star IV-203 probably isn't a problem star at all.  The spread in 
\teff{} results is large, but that might be due to the increased 
line blanketing in this cool star. In this
case, the standard deviation between the three \teff{} method results 
is 59 K.  This is larger than most of the other stars, but we believe this
is not enough to merit any action.

For II-122 and IV-025 the initial analysis showed large differences 
in the ionization balance between Fe~I and Fe~II 
($\log\epsilon{\rm (FeII)}-\log\epsilon{\rm (FeI)}=+0.43$ for II-122 
and $-0.53$ for
IV-025).  The values of \tphot{} and \tex{} show reasonable agreement,
while \tion{} is forced to change greatly to bring the star into 
ionization equilibrium.  The problem with these two stars is that we may 
have assumed the incorrect distance.  Therefore, we adjust the distance 
modulus from 14.51~mag to minimize
the ionization difference.  For IV-025, we adjust M$_V$ from $+0.33$
to $+2.05$, or to a new distance modulus of 12.78 (d~$= 3.6$~kpc), while
for II-122 the M$_V$ change was from $-1.57$ to $-2.92$ (a new distance 
modulus of 15.89, d~$= 15.1$~kpc).  At these new distances, the ionization
differences are reduced to less than 0.05~dex and the \teff{} indicators
agree to within about 120~K.

The need for a change in distance scale is further demonstrated in Figure~6. 
In Figure~6(a), we show the excitation plots (\leps(Fe) vs. Excitation
Potential) for the three methods for IV-025 assuming a distance of 8~kpc.  
For the
photometric and excitation \teff{} methods, the slope of the plot is
very flat, but the Fe~II abundances are much below the Fe~I abundances.
The \teff{} change needed to bring the two species into agreement 
greatly affects the slope of the plot.

In Figure~6(b), we plot the same data except that the M$_V$ value
of IV-025 has been changed to the final value of $+2.05$.  In all three 
panels the slope of the Fe~I lines is flat and the mean abundance of the 
Fe~II lines agree with the mean of the Fe~I lines.  

The new distances to these two stars places them outside the bulge.  The
low metallicity of II-122, at [Fe/H]~$=-0.79$, and its distance, approximately 
7~kpc beyond the bulge and over 1~kpc away from the plane of the disk,
suggests that it might reasonably be a member of the thick-disk or halo.
Star IV-025 is metal-rich, at [Fe/H]~$=+0.21$, but it is about 250~pc above
the plane of the disk, roughly half way between the Sun and the bulge.
We will continue to include these stars in our analysis, but will not include 
them in any future discussion of the bulge and its abundance properties.

\subsection{Final Abundances}

A summary of the \teff{} and iron abundance results from the three methods
is given in Table~8.  The results for II-122 and IV-025 reflect the new
assumed distances. 
Since most of the \teff{} values given by the three methods are in
very good agreement, we have decided to adopt the mean of the three
(two in the case of IV-027, II-172 and I-322, as noted above) individual 
\teff{} values as our final adopted \teff.  The analyses were re-run using
these new \teff{} values; the final parameters and abundance results
are given in Table~9.  In Table~9 and elsewhere, we use [m/H] to denote 
the scaled solar metallicity used in the stellar atmosphere model, and
[Fe/H] to signify the iron abundance as derived from our analysis on our
scale where the solar \leps(Fe) = 7.45.

\subsection{Comparison to Previous Results}

In Figure~7 we compare the [Fe/H] values derived in this paper (``[Fe/H] New'')
against those derived by Rich (1988; R88), McWilliam \& Rich (1994; MR94),
Sadler, Rich \& Terndrup (1996; SRT96) and McWilliam \& Rich (2004; MR04).
For the R88 data, we fit to the ``Solution~1'' values.
We do the comparison against all stars that match between the samples,
including the non-bulge stars II-122 and IV-025, but excluding any solar 
neighborhood stars shared in the samples (three stars in R88, two in MR94).  
In all four cases, the metallicities derived here are slightly lower than
derived in previous works, especially at the metal-poor end.  
The mean differences between the common stars (in the sense of ``old minus 
new'') are $+0.34 \pm 0.18$~dex for the R88 sample, $+0.04 \pm 0.17$~dex for
the MR94 sample, $+0.12 \pm 0.54$~dex for the SRT96 sample and 
$+0.04 \pm 0.10$~dex for the MR04 sample.  

The fits were performed using the ``least-squares cubic'' method of 
\citet{y66}, which does not assume that the independent variables (here 
taken to be our new [Fe/H] values) are free of uncertainty.  For the 
fits to the R88 sample, we use the $\sigma$[Fe/H] value given in their 
Table~11.  That estimate is explicitly for their Solution~1, but we have used
that value for all three solutions because no other values are available.
For the SRT96 sample we use their $\sigma$[Fe/H]$ = 0.24$ estimate.  
For the MR94, we use their $\sigma$[Fe/H] value.  Finally, for the MR04 
sample, we give both our and their [Fe/H] values equal weight since both 
analyses came from the same data:

\begin{equation}
{\rm [Fe/H]}_{\rm R88} = 0.907 {\rm [Fe/H]}_{\rm New} + 0.311; \sigma = 0.173;
r = 0.956; N = 21
\end{equation}

\begin{equation}
{\rm [Fe/H]}_{\rm MR94} = 0.802 {\rm [Fe/H]}_{\rm New} - 0.025; \sigma = 0.131;
r = 0.968; N = 11
\end{equation}

\begin{equation}
{\rm [Fe/H]}_{\rm SRT96} = 1.027 {\rm [Fe/H]}_{\rm New} + 0.123; \sigma = 0.564;
r = 0.669; N = 17
\end{equation}

\begin{equation}
{\rm [Fe/H]}_{\rm MR04} = 0.938 {\rm [Fe/H]}_{\rm New} + 0.012; \sigma = 0.099;
r = 0.989; N = 9
\end{equation}
The value of $\sigma$ in the above equations indicates the formal root mean 
square scatter about each fit, and does not include the uncertainties on the 
individual [Fe/H] measurements from any of the works.  The value of $r$ is
the correlation coefficient, and $N$ is the number of starts matched between
the two samples.  

The solution for the SRT96 data includes two extreme outliers (III-220 and 
IV-047).  Our calculated abundances for each star is over 1~dex different than
what was derived by SRT96 (neither star in in R88).  Such a difference could
be due to mis-identifications by SRT96 or problems with their analysis 
procedure.  SRT96 assigns very high metallicities ([Fe/H] $> +1$) to some
Baade's Window giants, yet we do not confirm this result with our analysis.
Three of our stars (I-025, I-039 and IV-047) were given [Fe/H] values larger
than $+0.8$ by SRT96, but we find that all of them have [Fe/H] $\leq +0.51$.
If the two extreme outliers are excluded from the comparison, the new fit 
becomes:
\begin{equation}
{\rm [Fe/H]}_{\rm SRT96} = 0.934 {\rm [Fe/H]}_{\rm New} + 0.089; \sigma = 0.258;r = 0.906; N = 15.
\end{equation}
We will use this second calibration between SRT96 and our results for the
the derivation of the bulge metallicity distribution function in the next
section.  Note that the mean [Fe/H] difference between the 15 remaining 
common stars between the SRT96 and present sample becomes $+0.11 \pm 0.25$~dex.
The removal of the outlyers barely changes the value of the mean but greatly
reduces the size of the standard deviation.

\subsection{The Bulge Metallicity Distribution Function}

The metallicity distribution function (MDF) is a powerful tool for
understanding the star formation history within a population \citep{s59,ss72}.  
\citet{r90} measured the first MDF for the bulge and found that it was well fit 
by the classic closed box gas exhaustion model.  This result means that the
bulge does not suffer from the ``G-dwarf'' problem seen in the local disk.
In practice, the wide abundance range was known since the 1950's,
as Baade and others knew that RR~Lyrae stars (indicative of the old metal-poor
population) and M-giants (indicative of the disk-like population) were present
in the same field.  This study confirms the presence of this wide abundance
range.

While our sample size is not large enough to create an MDF from the 
high-resolution sample alone, we can use the fits derived above to 
recalibrate the larger low-resolution spectroscopic samples of R88 and SRT96
using the least-squares fits from the previous section.
After recalibration, we calculate the value of the mass fraction of 
metals (Z) assuming Z$_{\odot} = 0.019$ and scaled solar element ratios.
From MR94, RM00 and MR04, we know that certain $\alpha$-elements such as 
Mg are enhanced
from solar, so the values of Z derived here are too low.  It
will be necessary to revisit the MDF once we have derived the abundances
of other elements, especially the common heavy elements O, C, N, Mg, and Si.

Care has to be taken in selecting samples when calculating an MDF. 
In the case of the bulge, there is contamination by foreground and
background stars.  More importantly, very metal-rich giants quickly
evolve into cool, strong-lined M-giants.  Metallicity determination, either
by high- or low-resolution methods, become very problematic in M-giants.
Worse, near the boundary between K-giants and M-giants a slight increase in the 
[Ti/Fe] ratio could be enough to form the strong TiO bands characteristic 
of M-giants, but at higher \teff{} than for solar neighborhood stars.  This
could create a sampling bias in which a K-giant sample under-estimates
the frequency of stars with high [Ti/Fe] ratios.  

Our metallicity recalibration is based on K-giants, so the most desirable
situation is to use stars on a part of the giant branch where stars of all
abundances are K-giants.  Fortunately, SRT96 identified a subsample of 217 
stars at and below the RGB clump that belong to the bulge population (196
clump stars and 21 bulge giants with V$ > 17$).  
These stars are warm enough (\teff{}~$> 4000$, as determined by SRT96) that
even the coolest cannot have significant TiO formation.  Many of these SRT96
stars (including stars 
we observed) have TiO band measures from R88.  None of the SRT96 bulge clump 
stars that were observed in R88 have TiO band strength greater than the values
observed in one or more of our recalibration stars.  Thus, by using our sample
of mostly clump stars 
to define the bulge MDF we count the stars without the metallicity and
composition bias' associated with stars that become M-giants.
Since all bulge red giant stars go through the He-burning (HB) red clump
it is an excellent region to fully sample the MDF.
We note that although the lifetimes of the helium burning phase
is affected by metallicity, with a longer HB lifetime for more metal-rich stars,
the size of the effect is relatively small: approximately 10\% increase in HB
lifetime for a 1~dex increase in metallicity 
(e.g. see Girardi et al. 2000; Cassisi et al. 2004).  Our estimate of the 
correction to the MDF due to this effect, based on the tracks of \citet{gir00},
is similar to the uncertainty on the measured mean metallicity, with the 
corrected mean metallicity lower by $\sim$0.02--0.03~dex.

An additional advantage of using
clump stars is that the region of the color-magnitude diagram where they lie
is dominated by bulge stars (P(bulge) $> 80 \%$, as discussed in Section~2).
Therefore, we have chosen to use only the 217 stars identified by SRT96
as bulge clump stars (out of the 268 stars they identify as bulge members)
to define the bulge MDF.  In a separate MDF estimate we will use the full
sample of 88 stars from R88 because the initial star selection for that work
focused on K-giants.

A final caveat: the recalibration performed here cannot correct for any
selection bias or incompleteness in the initial samples.  The reader should
refer to the original papers for more detailed information.  The advantage of
our recalibration is that the high-resolution work, performed here, provides a
way to correct for zero-point and scale errors of the SRT96 and R88
low-resolution metallicity scales.

In Figures 8 and 9 we plot the recalibrated MDF for the R88 and SRT96
data. Also plotted are closed-box gas exhaustion models \citep{r90} using
either the mean or median value of Z as the yield.  Use of the median Z 
value in the model reduces the possibility that errors at either end of the 
recalibration offsets the final yield.  The mean [Fe/H] values
of the three samples are:   $-0.25\pm0.06$ (we quote the 
standard deviation of the mean for these mean values) for R88 Solution~1, 
$-0.15\pm0.03$ for the 268 bulge stars in the full SRT96 sample,
$-0.10\pm0.04$ for the 217 clump plus faint bulge giants and
$-0.07\pm0.04$ for the 196 clump-only stars.
The median values for these four samples are $-0.20$ for the R88 sample,
and $-0.13$, $-0.04$ and $-0.01$ for the three SRT96 subsamples, respectively.
If Equation~8 is used for the calibration of the SRT96 sample 
instead of Equation~10, the mean and median [Fe/H] for the 217 clump plus faint 
bulge giant stars become $-0.12\pm0.03$ and $-0.07$.

The difference in the mean and median Z value for R88 is small compared
to the scatter of the least-squares fit.  The closed box model fits the R88 
MDF well until approximately solar metallicity.  The model under-predicts
the number of stars with metallicities slightly greater than solar, then
slightly over-predicts the number of very metal-rich stars.  

\citet{z03} gets a similar result for their fits to their MDF based on
a M$_{\rm K}$ vs. (V$-$K)$_0$ CMD of Baade's Window.  They admit that the
calibration of the fiducial RGB shapes they use are dependent on the 
observed abundances of the metal-rich globular clusters NGC~6528 and NGC~6553.  
They also find a sizable deficit of metal-poor stars (Z~$<$~0.4~Z$_{\odot}$)
as compared to their model with a yield of solar metallicity.  

The MDF from the recalibrated SRT96 clump star data shows a larger difference 
between the mean and median values of Z than the R88 sample, but this is
reduced when compared to what is seen in the full-sample recalibration.  
For the full 268 star SRT96 recalibration, the mean Z value is 
1.36~Z$_{\odot}$, while the median Z is 0.75~Z$_{\odot}$, as compared to
1.54 and 0.91~Z$_{\odot}$, respectively, for the clump subsample.  Both SRT96
samples contain many more super-solar metallicity stars than R88.

From Figure 9, it appears that a yield between the mean and median may be 
more appropriate.  A higher yield
fits the high-metallicity end of the distribution better, although it slightly
under-predicts the number of metal-poor stars.  A lower yield over-predicts the
number of metal-poor stars and under-predicts the metal-rich end as well.  

\subsection{Bulge and Disk Metallicity Distribution Functions}

In Figure~10 we plot the MDF from the corrected SRT96 clump sample to represent
the bulge, and the MDF from \citet{ap04} to represent the disk.  \citet{ap04} analyzed 
high-resolution spectra of the 104 stars within 15~pc of the sun.  
The internal errors of the \citet{ap04} data are much lower than the
recalibrated SRT96 data, so we convolved the MDF 
with the uncertainty of the SRT96 data. 
This was done using by a summation of 104 Gaussians with $\sigma = 0.283$~dex,
each centered the [Fe/H] value of one of the stars in the \citet{ap04} sample.
The summation was then binned and normalized.  

The distribution of bulge stars is wider than the disk stars.  For the 
\citet{ap04} MDF to have the same width as the bulge MDF, the width of
the Gaussians used in the convolution above would have to be increased
to about 0.45~dex, which is much larger than the measurement uncertainties
of either work.  The mean value
of the raw disk sample is $-0.13 \pm 0.03$ (standard deviation of the mean), or 
slightly below the mean of the SRT96 clump sample.  Despite having a slightly
higher
mean value, the normalized bulge distribution has more very metal-poor stars
than the disk sample (clear in the most metal-poor bin in the lower panel
of Figure~10).  We confirm R88:  the ``G-dwarf'' problem, if it exists 
at all in the bulge, is less severe than in the local disk.  

Solutions of the ``G-dwarf'' problem in the local disk include such 
theories as the infall of primordial gas \citep{l74} or pre-enrichment of
gas by an earlier star formation \citep{ot75}.  The lack of a ``G-dwarf''
problem in the bulge would limit the amount of infall or pre-enrichment
needed to form the bulge.  However, our sample size is small (only 12 stars 
with recalibrated [Fe/H] values below $-1$; the closed-box model predicts 
13.6 stars within this metallicity range), and complexities such as the
shorter lifetimes of metal-poor HB stars and selection biases must be
taken into consideration.  If a number of these stars were not bulge
members our conclusion could change.  High resolution observations of all the
metal-poor Baade's Window stars would improve the derived metallicities and 
memberships of these stars.

Although the absence of a ``G-dwarf'' problem in the Galactic bulge complicates
comparisons of its MDF with that of the disk, it is pertinent to ask
how the bulge mean metallicity fits with the radial metallicity gradient
seen in the Galactic disk (e.g. Janes 1979; Twarog et al. 1997).  Given the
mean [Fe/H]$ = -0.13$~dex for solar neighborhood stars within 15~pc of 
the Sun from Allende Prieto et al. (2004), and the best estimate of the 
radial [Fe/H] gradient, of $-0.07$~dex per~kpc, from Twarog et al. (1997), 
the mean [Fe/H] of the disk at the Galactic bulge is expected to be 
$\sim +0.45$~dex.  
This value is significantly higher than our recalibration of the SRT96 dataset,
with the mean [Fe/H]~$= -0.10 \pm 0.04$~dex; indeed the extrapolated mean disk
value is roughly as high as the most metal-rich star in our sample of 27 bulge 
K~giants.  
Although it is possible that the inner disk does have a mean metallicity of 
[Fe/H]~$\sim +0.45$~dex, with the bulge significantly lower, we favor the 
idea that, approximately within the solar circle, there is either no radial
[Fe/H] gradient, or a very shallow gradient.  This conclusion is the same as
suggested by MR94 and Twarog et al. (1997); even Janes (1979) conceded the
possibility of a zero gradient in the inner disk.  The conclusion is based on
the lack of a clear metallicity gradient in the range 6.5~$\leq$~R~$\leq$10~kpc,
and the absence of data for galactocentric radii less than 
R$_{\rm G} \sim 6$~kpc; outside R$_{\rm G} \sim 8$--10~kpc a strong 
metallicity gradient is clear in the extant data.

\section{Summary}

We have performed detailed abundance analysis of 27 K-giants in Baade's Window,
together with a control sample of 17 giants from the solar neighborhood.  In
this paper we focus on the stellar atmosphere parameters and iron abundances.

Because of the difficulties associated with the high interstellar extinction 
toward Baade's Window we have employed three techniques to determine stellar 
effective temperature: photometric \teff{}, based on V$-$K colors and the 
color-temperature calibration of \citet{a99}; differential excitation temperatures, based 
on excitation equilibrium from Fe~I lines; and differential ionization temperatures, based 
on ionization equilibrium of Fe~I and Fe~II lines.  
The differential temperature estimates were taken relative to
the bright K1.5 giant Arcturus, whose effective temperature is known to high accuracy.
This differential technique offers the advantage that it is not limited by the 
paucity of Fe lines with accurate laboratory \gf-values that are useful for analysis of
red giant stars, and avoids the use of a disjoint collection of \gf-values with
a mixture of random and systematic uncertainties.  Furthermore, with our differential
abundances we expect some degree of canceling of abundance errors resulting from
processes not included in the construction of the model atmosphere grid.

Our analysis of the 17 
solar neighborhood giants confirms the consistency of the three methods used 
to estimate \teff{} when reddening is negligibly small.
Among the heavily reddened Baade's Window stars these three techniques for temperature 
estimation have allowed us to identify two of our stars as non-bulge members.

In our abundance analyses we have included spectrum synthesis to identify lines
blended with CN rotation-vibration lines, and random atomic features from the 
Kurucz database.  In this way we have produced a list of iron lines that are 
clean in the metal-rich giant $\mu$~Leo.  Strengths of these iron lines are 
suitable for abundance analysis in both $\mu$~Leo and Arcturus, which cover a 
range of $\sim$0.8~dex in metallicity.

We have also performed spectrum synthesis calculations to identify clean 
continuum regions in our stars.  For heavily blanketed stars, where there are 
insufficient clean continuum regions, we estimated the blanketed flux in 
portions of the spectrum closest to the expected continuum.  Typically we 
consider ``dirty'' continuum regions with less than 5\% blanketing.  Our 
syntheses provide us with flux corrections for these dirty
continuum regions, to apply to the observed spectrum when making the continuum 
normalization.  These continuum regions were especially important in the 
analysis of the metal-rich bulge stars, with the greatest line blanketing.  
We found that one result of misplaced continuum levels is large errors 
in the derived excitation temperatures.

Both the iron lines and continuum regions employed in this work are of general 
utility for the measurement of K-giant spectra with [Fe/H] values between
about $-1.5$ and $+0.5$.

The differential abundances of our bulge stars were put onto the solar scale by
computing the differential abundances of our clean lines in Arcturus relative to
the sun, based on EW measurements of those lines in the Kurucz et al. (1984) solar
spectrum and the Hinkle et al. (2000) spectrum of Arcturus.   In this way we find
[Fe/H]=$-$0.50 for Arcturus, with a scatter of 0.07~dex.  The random error on the 
mean is less than 0.01~dex, but we believe that the total uncertainty is dominated by 
systematic effects, which we estimate may be $\sim$0.05~dex.

Our high-resolution iron abundance results agree closely with the values given 
for stars common to the samples of MR94 and MR04.  The [Fe/H] values in our 
high-resolution sample of bulge stars ranges from $-1.29$ to $+0.51$~dex, 
with a mean of $-0.37$~dex; although the mean has no significance beyond a 
reflection of our sample selection biases.
The mean differences between the common stars (in the sense of ``old minus 
new'') are $+0.34 \pm 0.18$~dex for the R88 sample, $+0.04 \pm 0.17$~dex for
the MR94 sample, $+0.12 \pm 0.54$~dex for the SRT96 sample and 
$+0.04 \pm 0.10$~dex for the MR04 sample.  

In order to estimate the bulge metallicity distribution function we have 
corrected the results of R88 and SRT96, which were based on low-resolution 
spectra, onto our system by transformation equations based on stars in common 
with the current investigation.

For the transformed R88 sample (Solution~1) we obtain a mean 
[Fe/H]$= -0.25 \pm 0.06$~dex, consistent with the re-calibration of the R88 
work by MR94.  For the re-calibrated full sample of SRT96 we obtain a mean 
[Fe/H] $= -0.15 \pm 0.03$~dex.

The stars studied here and in SRT96 and R88, were restricted to K-giant stars, 
and thus give a biased metallicity distribution function.  To avoid this bias 
we have selected SRT96 stars identified by them as members of the bulge red 
clump.  Because all red giant stars in the bulge pass through the red clump 
phase, the red clump stars offer the potential for the most reliable estimate 
of the true bulge metallicity function for giant stars.  We estimate that a 
small bias, due to slightly longer lifetimes for metal-rich red clump stars 
than the metal-poor variety requires a decrease of approximately 0.02 to 
0.03~dex in the mean [Fe/H] of the metallicity distribution function.  
Thus, our best estimate for the mean [Fe/H] of the Galactic bulge is 
$-0.10 \pm 0.04$~dex, based on our high-resolution re-calibration of the
SRT96 results.

Our re-calibration of the SRT96 data, together with abundance results for solar 
neighborhood stars by Allende Prieto et al. (2004), confirms the conclusion, 
first made by Rich (1990), that the G-dwarf problem (i.e. a deficit of 
metal-poor stars relative to the Simple chemical evolution model) is either 
non-existent in the bulge, or much less severe than in the solar neighborhood.

Although the bulge and inner disk are two separate systems we note that the 
mean bulge [Fe/H] measured here, at $-0.10$~dex, lies far below the 
extrapolated mean [Fe/H] value for the inner Galactic disk, at 
$\sim +0.45$~dex, based on a linear Galactic radial metallicity gradient of 
$-0.07$~dex per~kpc (e.g. Twarog et al. 1997).  Instead, we favor the idea of
a zero, or much reduced, radial metallicity gradient in the Galactic
disk, for R$_{\rm G}$ within the solar circle, as discussed by MR94.

\acknowledgments

We are especially grateful to the staff of Keck 
Observatory for their assistance, and S. Vogt and his team for building 
HIRES.  The authors would like to thank Ruth Peterson for many useful
suggestions in the referee report.  We acknowledge support from 
grant AST-0098612 from the 
National Science Foundation.  For the Arcturus abundance analysis we gratefully
acknowledge partial support from a NASA-SIM Key Project grant, entitled
{\it ``Anchoring the Population II Distance Scale: Accurate Ages for Globular Clusters
and Field Halo Stars''}.
  The authors acknowledge the cultural role that
the summit of Mauna Kea has had within the indigenous Hawaiian community.
We are fortunate to have the opportunity to conduct observations from this
mountain.  This publication makes use of data products from the Two Micron 
All Sky Survey, which is a joint project of the University of Massachusetts 
and the Infrared Processing and Analysis Center/California Institute of 
Technology, funded by the National Aeronautics and Space Administration 
and the National Science Foundation.  This research has also made use of the
SIMBAD database, operated at CDS, Strasbourg, France.

\begin{deluxetable}{rrrcrrccc}
\tabletypesize{\footnotesize}
\tablenum{1}
\tablewidth{0pt}
\tablecaption{Keck/HIRES Journal of Observations}
\tablehead{
\colhead{Name} & \colhead{V$_{\circ}$} & \colhead{(V$-$K)$_{\circ}$} & \colhead{Date} & \colhead{Slit\tablenotemark{a}} & \colhead{Exp.} & \colhead{Coverage} & \colhead{S/N} & \colhead{E(B$-$V)} \\
\colhead{} & \colhead{} & \colhead{} & \colhead{} & \colhead{} & \colhead {(s)} & \colhead{(\AA)} &\colhead{per pixel\tablenotemark{b}} &\colhead{(mag)} 
}
\startdata
  I-012 & 14.37 & 2.929 & 2000-07-04 &C1&  6000 &  5425--7750 &  90 & 0.430 \\
  I-025 & 15.42 & 2.864 & 2000-07-03 &B2&  8000 &  5425--7750 &  45 & 0.427 \\
  I-039 & 15.59 & 2.723 & 1998-08-16 &C1&  6000 &  5475--7825 &  55 & 0.414 \\
  I-141 & 14.55 & 2.907 & 2001-07-16 &B2&  6000 &  5500--7900 &  75 & 0.420 \\
  I-151 & 14.47 & 2.708 & 2001-08-14 &C1&  6000 &  5500--7900 &  60 & 0.423 \\
  I-152 & 15.42 & 2.431 & 2001-08-14 &C1&  6000 &  5500--7900 &  45 & 0.423 \\
  I-156 & 14.75 & 2.908 & 2001-08-14 &C1&  6000 &  5500--7900 &  60 & 0.423 \\
  I-158 & 15.15 & 2.828 & 2001-08-14 &C1&  6000 &  5500--7900 &  55 & 0.423 \\
  I-194 & 14.77 & 3.134 & 1998-08-15 &C1&  8000 &  5500--7825 &  75 & 0.426 \\
  I-202 & 14.46 & 3.050 & 1999-09-01 &C1&  7500 &  5425--7875 &  70 & 0.426 \\
  I-264 & 12.97 & 3.269 & 2000-08-01 &C1&  1800 &  5375--7750 &  85 & 0.447 \\
  I-322 & 13.01 & 3.131 & 1999-09-09 &C1&  4500 &  5425--7875 & \llap{1}30 & 0.463 \\
 II-033 & 13.97 & 2.890 & 2000-08-01 &C1&  3600 &  5375--7750 &  75 & 0.477 \\
 II-119 & 14.14 & 2.552 & 2000-08-01 &C1&  3600 &  5375--7750 &  60 & 0.488 \\
 II-122 & 12.98 & 3.769 & 2001-06-28 &C1&  4000 &  5475--7750 &  80 & 0.502 \\
 II-154 & 14.97 & 2.440 & 2000-08-02 &C1&  6300 &  5375--7750 &  55 & 0.500 \\
 II-172 & 15.46 & 2.958 & 2000-08-02 &C1&  7500 &  5375--7900 &  55 & 0.490 \\
III-152 & 14.64 & 3.124 & 1999-08-18 &C1&  4500 &  5500--7875 &  85 & 0.438 \\
III-220 & 14.97 & 2.571 & 2001-07-16 &C1&  6000 &  5425--7875 &  60 & 0.515 \\
 IV-003 & 13.70 & 2.711 & 1999-07-19 &C1&  4500 &  5425--7875 &  80 & 0.439 \\
 IV-025 & 14.84 & 2.419 & 1999-07-18 &C1&  8000 &  5500--7875 &  40 & 0.438 \\
 IV-047 & 15.68 & 1.07\tablenotemark{c}& 2000-07-03 &B2&  6000 &  5425--7750 &  40 & 0.414 \\
 IV-072 & 14.93 & 2.984 & 1998-08-14 &C1& 14000 &  5425--7850 & \llap{1}00 & 0.428 \\
 IV-167 & 15.59 & 2.946 & 1999-07-19 &B2& 10000 &  5425--7875 &  75 & 0.450 \\
 IV-203 & 12.52 & 3.776 & 1999-09-11 &C1&  1440 &  5425--7875 &  70 & 0.435 \\
 IV-325 & 15.58 & 2.958 & 2000-08-01 &C1&  3600 &  5350--7875 &  70 & 0.414 \\
 IV-329 & 13.81 & 3.103 & 2000-04-20 &C1&  4200 &  5425--7850 &  90 & 0.414 \\
 $\mu$ Leo &3.88&2.66&2000-07-04&B2&2&5425--7750 & \llap{2}20 & 0.000 \\
\enddata
\tablenotetext{a}{Slit C1 is 0.861 arcsec by 7 arcsec and yields an 
instrumentalresolution of 45000, while slit B2 is 0.571 arcsec by 7 arcsec
and yields a resolution of 60000. }
\tablenotetext{b}{Peak value in the order containing H$\alpha$.}
\tablenotetext{c}{Value given for IV-047 is (B$-$V)$_{\circ}$ based on Arp (1965) photometry.  The V$_{\circ}$ is based on TSR95 data.}
\end{deluxetable}

\begin{deluxetable}{lcccccc}
\tabletypesize{\footnotesize}
\tablenum{2}
\tablewidth{0pt}
\tablecaption{Las Campanas/Lick Journal of Observations\tablenotemark{a}}
\tablehead{
\colhead{Name} & \colhead{V$_{\circ}$} & \colhead{(B$-$V)$_{\circ}$} & \colhead{(V$-$K)$_{\circ}$} & \colhead{Date} & \colhead{Exp.} & \colhead{S/N} \\
\colhead{} & \colhead{} & \colhead{} & \colhead{} & \colhead {} & \colhead{(sec)} &\colhead{per pixel} 
}
\startdata
HR1184   $\rho$ For     & 5.54 & 0.98& \nodata&  2003-12-31 & \llap{1}5 & 175 \\
HR1346   $\gamma$ Tau   & 3.65 & 0.98&  2.14  &  2003-12-31 &  3 & 175 \\
HR1348   $\phi$ Tau     & 4.96 & 1.15& \nodata&  2003-12-31 & \llap{1}0 & 200 \\
HR1409   $\epsilon$ Tau & 3.54 & 1.03&  2.21  &  2003-12-31 &  3 & 175 \\
HR1411   $\theta$ Tau   & 3.85 & 0.95&  2.10  &  2003-12-31 &  3 & 140 \\
HR1585                  & 5.50 & 1.31& \nodata&  2003-12-31 & \llap{1}0 & 150 \\
HR2035   $\delta$ Lep   & 3.81 & 0.99&  2.54  &  2004-01-02 &  3 & 200 \\
HR2113\tablenotemark{b} & 4.37 & 1.17&  2.83  &  2004-01-06 &  5 & 220 \\
HR2443   $\nu$ Cma      & 4.43 & 1.15&  2.61  &  2004-01-06 &  5 & 210 \\
HR3418   $\sigma$ Hya   & 4.45 & 1.22&  2.64  &  2004-01-06 &  5 & 210 \\
HR3733   $\lambda$ Pyx  & 4.73 & 0.91&  2.05  &  2003-12-31 &  5 & 190 \\
HR4104   $\alpha$ Ant   & 4.28 & 1.45&  3.41  &  2003-12-31 &  5 & 270 \\
HR4382   $\delta$ Crt   & 3.56 & 1.11&  2.61  &  2004-01-05 &  5 & 325 \\
HR4450   $\xi$ Hya      & 3.54 & 0.96&  2.09  &  2004-01-01 &  5 & 110 \\
HR4608   {\it o} Vir    & 4.13 & 0.97&  2.22  &  2004-01-03 &  5 & 240 \\
HR5340   $\alpha$ Boo\quad\quad   &\llap{$-$}0.04&1.23&  3.00  &  1997-05-13 & \llap{1}0 & 220 \\
\enddata
\tablenotetext{a}{All of the above observations were taken with the echelle
spectrograph on the du Pont 2.5-m telescope at Las Campanas Observatory
with the exception of HR5340 ($\alpha$ Boo).  The spectrum of $\alpha$ Boo
was taken with the 0.6-m CAT telescope at Lick Observatory and the Hamilton
spectrograph.  See the text for more details.}
\tablenotetext{b}{Assumes A$_{\rm V} = 0.17$ from McWilliam (1990). }
\end{deluxetable}

\begin{deluxetable}{rrrccccc}
\tabletypesize{\footnotesize}
\rotate
\tablenum{7}
\tablewidth{0pt}
\tablecaption{Mean Effects from Changing Atmosphere Grids\tablenotemark{a}}
\tablehead{
\colhead{Atm. Grid} & \colhead{$\Delta$T$_{\rm Ex}$} & \colhead{$\Delta$T$_{\rm Ion}$} & \colhead{$\Delta\epsilon$(Fe I)$_{\rm Phot}$\tablenotemark{b}} & \colhead{$\Delta$(II-I)$_{\rm Ex}$\tablenotemark{c}}  & \colhead{$\Delta\epsilon$(Fe I)$_{\rm Ex}$} & \colhead{$\Delta$(I-II)$_{\rm Ex}$} & \colhead{$\Delta\epsilon$(Fe I)$_{\rm Ion}$}
}
\startdata
\multicolumn{8}{c}{ALL STARS}\\
ODFNEW  & $ -2$&$-15$&$-0.01$&$-0.02$&$+0.00$&$-0.01$&$-0.01$\\
$\sigma$& $+11$&$+21$&$+0.02$&$+0.03$&$+0.02$&$+0.02$&$+0.02$\\
AODFNEW & $ -3$&$-26$&$-0.01$&$-0.03$&$+0.00$&$-0.02$&$+0.00$\\
$\sigma$& $+24$&$+30$&$+0.02$&$+0.03$&$+0.01$&$+0.02$&$+0.01$\\
NOVER   & $ +3$&$ -8$&$-0.01$&$-0.01$&$+0.00$&$+0.01$&$-0.01$\\
$\sigma$& $+22$&$+12$&$+0.01$&$+0.01$&$+0.04$&$+0.06$&$+0.01$\\
MARCS   & $ +3$&$ +1$&$+0.00$&$+0.01$&$+0.01$&$+0.07$&$+0.00$\\
$\sigma$& $ 16$&$+26$&$+0.01$&$+0.04$&$+0.02$&$+0.10$&$+0.02$\\
\multicolumn{8}{c}{STARS WITH [Fe/H] $> 0$}\\
ODFNEW  & $-12$&$-38$&$-0.03$&$-0.04$&$-0.02$&$-0.02$&$-0.03$\\
$\sigma$& $+14$&$+12$&$+0.02$&$+0.03$&$+0.02$&$+0.02$&$+0.01$\\
AODFNEW & $-26$&$-59$&$-0.01$&$-0.07$&$+0.01$&$-0.02$&$+0.00$\\
$\sigma$& $+24$&$+17$&$+0.01$&$+0.02$&$+0.01$&$+0.02$&$+0.01$\\
NOVER   & $ +3$&$-13$&$+0.00$&$-0.01$&$+0.00$&$-0.01$&$+0.00$\\
$\sigma$& $ +7$&$ +8$&$+0.01$&$+0.01$&$+0.01$&$+0.02$&$+0.01$\\
MARCS   & $+22$&$+28$&$+0.01$&$+0.04$&$+0.02$&$+0.08$&$+0.01$\\
$\sigma$& $ 13$&$+23$&$+0.01$&$+0.03$&$+0.01$&$+0.04$&$+0.02$\\
\enddata
\tablenotetext{a}{Differences defined as the result from the listed
atmosphere minus the result from the default Kurucz atmosphere
(solar abundance ratios with overshoot).}
\tablenotetext{b}{For shorthand, we will use $\epsilon$ instead of $\log{\epsilon}$ in table headers.}
\tablenotetext{c}{The ionization difference is defined as the value
of $\log{\rm \epsilon(Fe II)} - \log{\rm \epsilon(Fe I)}$.}
\end{deluxetable}

\begin{deluxetable}{rrrrrrrrrrrrr}
\tabletypesize{\footnotesize}
\rotate
\tablenum{8}
\tablewidth{0pt}
\tablecaption{Parameter Determinations}
\tablehead{
\colhead{Name} & \colhead{M$_{\rm V}$} & \colhead{Mass} & \colhead{T$_{\rm Phot}$} & \colhead{$\epsilon$(FeI)} & \colhead{$\Delta$(II-I)} &\colhead{T$_{\rm Ex}$} & \colhead{$\epsilon$(FeI)} & \colhead{$\Delta$(II-I)} &\colhead{T$_{\rm Ion}$} & \colhead{$\epsilon$(FeI)} & \colhead{$<$T$>$} & \colhead{$\sigma$T} 
}
\startdata
\multicolumn{13}{c}{BAADE'S WINDOW BULGE STARS} \\
I-012  &$-0.14$&0.8&4278&7.08&$+0.04$&4204&7.06&$+0.09$&4289&7.08&4257&46\\
I-025  &$+0.90$&0.8&4331&7.97&$+0.05$&4324&7.96&$+0.05$&4364&7.95&4340&21\\
I-039  &$+1.09$&0.8&4430&7.96&$-0.09$&4371&7.94&$-0.03$&4356&7.93&4386&39\\
I-141  &$+0.04$&0.8&4293&7.17&$+0.08$&4376&7.18&$-0.03$&4336&7.17&4335&42\\
I-151  &$-0.04$&0.8&4434&6.69&$-0.03$&4402&6.67&$-0.01$&4379&6.65&4405&28\\
I-152  &$+0.91$&0.8&4666&7.45&$-0.03$&4609&7.43&$+0.04$&4664&7.45&4646&32\\
I-156  &$+0.24$&0.8&4292&6.74&$+0.04$&4323&6.74&$-0.01$&4320&6.75&4312&17\\
I-158  &$+0.64$&0.8&4347&7.25&$-0.04$&4368&7.24&$-0.08$&4333&7.24&4349&18\\
I-194  &$+0.26$&0.8&4153&7.20&$ 0.00$&4218&7.20&$-0.10$&4157&7.19&4176&36\\
I-202  &$-0.05$&0.8&4204&7.61&$-0.05$&4184&7.59&$-0.08$&4163&7.60&4184&21\\
I-264  &$-1.54$&0.8&4081&6.29&$+0.05$&4103&6.29&$-0.01$&4108&6.29&4097&14\\
I-322  &$-1.50$&0.8&4155&7.20&$-0.13$&4293&7.23&$-0.37$&4058&7.20&4106&69\\
II-033 &$-0.54$&0.8&4304&6.71&$-0.04$&4279&6.70&$-0.01$&4247&6.68&4277&29\\
II-119 &$-0.37$&0.8&4563&6.24&$-0.04$&4605&6.28&$-0.10$&4493&6.18&4554&57\\
II-154 &$+0.46$&0.8&4657&6.85&$-0.02$&4662&6.85&$-0.02$&4630&6.83&4650&17\\
II-172 &$+0.95$&0.8&4260&7.11&$+0.33$&4447&7.14&$+0.10$&4514&7.16&4480&47\\
III-152&$+0.13$&0.8&4158&7.04&$ 0.00$&4165&7.03&$-0.02$&4149&7.02&4157& 8\\
III-220&$+0.46$&0.8&4543&7.13&$-0.01$&4565&7.14&$-0.03$&4542&7.13&4550&13\\
IV-003 &$-0.80$&0.8&4438&6.17&$-0.01$&4429&6.15&$+0.02$&4433&6.16&4433& 5\\
IV-047 &$+1.17$&0.8&4250&6.97&$+0.44$&4530&7.03&$+0.05$&4581&7.06&4556&36\\
IV-072 &$+0.42$&0.8&4250&7.71&$+0.05$&4300&7.68&$-0.06$&4266&7.68&4272&26\\
IV-167 &$+1.08$&0.8&4275&7.91&$+0.06$&4315&7.90&$-0.01$&4312&7.90&4301&22\\
IV-203 &$-2.00$&0.8&3856&6.15&$+0.23$&3882&6.15&$+0.17$&3969&6.16&3902&59\\
IV-325 &$+1.07$&0.8&4267&7.72&$+0.07$&4288&7.70&$+0.01$&4312&7.71&4289&23\\
IV-329 &$-0.70$&0.8&4171&6.50&$+0.08$&4205&6.50&$+0.01$&4216&6.51&4197&23\\
\multicolumn{13}{c}{BAADE'S WINDOW NON-BULGE STARS} \\
II-122\tablenotemark{a} &$-2.92$&0.8&3856&6.66&$+0.13$&3973&6.66&$-0.07$&3907&6.64&3919&59\\
IV-025\tablenotemark{b} &$+2.05$&0.8&4682&7.68&$-0.10$&4557&7.66&$+0.03$&4599&7.66&4514&66\\
\multicolumn{13}{c}{LOCAL DISK STARS} \\
HR1184 &$+0.85$&0.9&4765&7.07&$-0.02$&4750&7.07&$-0.01$&4743&7.06&4753&11\\
HR1346 &$+0.28$&1.9&4877&7.63&$-0.07$&4814&7.59&$-0.05$&4779&7.57&4823&50\\
HR1348 &$-0.14$&1.6&4452&7.10&$-0.04$&4364&7.07&$+0.06$&4412&7.08&4409&44\\
HR1409 &$+0.16$&2.6&4863&7.66&$+0.02$&4781&7.62&$+0.10$&4869&7.66&4838&49\\
HR1411 &$+0.43$&2.6&5013&7.65&$-0.02$&4891&7.59&$+0.11$&4979&7.64&4961&63\\
HR1585 &$-0.31$&1.0&4324&7.11&$-0.03$&4376&7.13&$-0.08$&4299&7.11&4333&39\\
HR2035 &$+1.13$&0.9&4661&6.83&$-0.04$&4595&6.79&$+0.02$&4617&6.80&4624&34\\
HR2113 &$-1.18$&1.3&4323&6.83&$-0.14$&4277&6.81&$-0.09$&4230&6.79&4277&46\\
HR2443 &$-1.34$&1.6&4432&7.24&$-0.05$&4460&7.25&$-0.10$&4395&7.22&4429&33\\
HR3418 &$-0.72$&2.2&4490&7.62&$+0.04$&4393&7.58&$+0.15$&4528&7.62&4470&70\\
HR3733 &$+1.00$&2.3&5040&7.36&$+0.04$&4945&7.29&$+0.09$&5017&7.33&5001&50\\
HR3905 &$+0.83$&1.3&4525&7.78&$+0.04$&4523&7.77&$+0.03$&4546&7.78&4531&13\\
HR4104 &$-0.97$&0.8&4045&7.09&$+0.04$&4057&7.06&$-0.05$&4037&7.05&4046&10\\
HR4382 &$-0.32$&1.0&4527&6.94&$-0.04$&4393&6.87&$+0.12$&4506&6.93&4475&72\\
HR4450 &$+0.55$&1.6&5011&7.60&$-0.07$&4996&7.59&$-0.06$&4917&7.53&4975&51\\
HR4608 &$+0.53$&1.6&4887&7.07&$-0.04$&4818&7.01&$+0.04$&4853&7.04&4853&35\\
HR5340 &$-0.20$&0.8&4285&6.95&$+0.04$&4244&6.93&$+0.06$&4321&6.96&4283&39\\
\enddata
\tablenotetext{a}{Adopted distance modulus of 15.89 mag, d~=~15.1~kpc.}
\tablenotetext{b}{Adopted distance modulus of 12.78 mag, d~=~3.6~kpc.}
\normalsize
\end{deluxetable}

\begin{deluxetable}{rrrrrrrrrrrrrrr}
\tabletypesize{\footnotesize}
\rotate
\tablenum{9}
\tablewidth{0pt}
\tablecaption{Final Parameters and Abundances}
\tablehead{
\colhead{Name} & \colhead{T$_{\rm eff}$} & \colhead{$\sigma$T} & \colhead{$\log{g}$} & \colhead{[m/H]\tablenotemark{a}} & \colhead{v$_{\rm t}$} &\colhead{$\sigma$v$_{\rm t}$} & \colhead{$\epsilon$(FeI)} & \colhead{$\sigma$$\epsilon$(FeI)} &\colhead{N(FeI)} &\colhead{$\epsilon$(FeII)} & \colhead{$\sigma$$\epsilon$(FeII)} &\colhead{N(FeII)} & \colhead{$\Delta$(II-I)} & \colhead{$\sigma\Delta$(II-I)} 
}
\startdata
\multicolumn{15}{c}{BAADE's WINDOW BULGE STARS} \\
I-012  &4257&46&1.55&$-0.37$&1.54&0.03&7.08&0.08&112&7.14&0.07&4&$+0.06$&0.11\\
I-025  &4340&21&2.02&$+0.51$&1.62&0.05&7.96&0.09& 90&8.00&0.07&5&$+0.04$&0.11\\
I-039  &4386&39&2.13&$+0.50$&1.47&0.06&7.95&0.09& 83&7.92&0.07&3&$-0.03$&0.11\\
I-141  &4335&42&1.68&$-0.27$&1.33&0.05&7.18&0.09& 99&7.20&0.03&4&$+0.02$&0.10\\
I-151  &4405&28&1.70&$-0.78$&1.22&0.06&6.67&0.08& 83&6.67&0.06&3&$-0.00$&0.10\\
I-152  &4646&32&2.20&$-0.01$&1.22&0.06&7.44&0.10& 78&7.44&0.06&3&$-0.00$&0.11\\
I-156  &4312&17&1.74&$-0.71$&1.16&0.05&6.74&0.07& 79&6.76&0.03&4&$+0.02$&0.08\\
I-158  &4349&18&1.93&$-0.20$&1.27&0.06&7.25&0.10&102&7.20&0.06&4&$-0.05$&0.11\\
I-194  &4176&36&1.65&$-0.25$&1.34&0.05&7.20&0.10& 96&7.17&0.09&3&$-0.03$&0.13\\
I-202  &4184&21&1.53&$+0.16$&1.12&0.04&7.61&0.11&107&7.58&0.09&4&$-0.03$&0.14\\
I-264  &4097&14&0.87&$-1.15$&1.67&0.10&6.30&0.08& 83&6.32&0.08&5&$+0.02$&0.12\\
I-322  &4106&69&0.89&$-0.25$&1.63&0.04&7.20&0.09&103&7.16&0.05&5&$-0.05$&0.11\\
II-033 &4277&29&1.41&$-0.75$&1.41&0.03&6.70&0.07& 92&6.71&0.03&4&$+0.01$&0.07\\
II-119 &4554&57&1.66&$-1.22$&1.24&0.13&6.23&0.10& 58&6.20&0.06&4&$-0.03$&0.11\\
II-154 &4650&17&2.03&$-0.61$&1.00&0.05&6.84&0.08& 83&6.84&0.07&5&$-0.00$&0.10\\
II-172 &4480&47&2.14&$-0.30$&1.03&0.05&7.15&0.11& 97&7.22&0.06&5&$+0.06$&0.13\\
III-152&4157& 8&1.58&$-0.41$&1.21&0.04&7.04&0.08&103&7.04&0.06&4&$+0.00$&0.10\\
III-220&4550&13&1.99&$-0.31$&1.27&0.01&7.14&0.08& 94&7.14&0.08&3&$+0.00$&0.11\\
IV-003 &4433& 5&1.41&$-1.29$&1.35&0.11&6.16&0.07& 58&6.16&0.07&4&$-0.00$&0.10\\
IV-047 &4556&36&2.28&$-0.40$&1.54&0.06&7.05&0.11& 85&7.10&0.04&4&$+0.05$&0.11\\
IV-072 &4272&26&1.78&$+0.26$&1.31&0.05&7.71&0.10& 91&7.72&0.02&3&$+0.02$&0.10\\
IV-167 &4301&22&2.07&$+0.46$&1.38&0.05&7.91&0.10& 89&7.94&0.05&4&$+0.03$&0.12\\
IV-203 &3902&59&0.51&$-1.29$&1.88&0.09&6.16&0.07& 80&6.32&0.03&4&$+0.16$&0.07\\
IV-325 &4289&23&2.05&$+0.28$&1.69&0.05&7.73&0.09& 95&7.77&0.05&5&$+0.04$&0.10\\
IV-329 &4197&23&1.29&$-0.94$&1.47&0.04&6.51&0.06& 77&6.55&0.04&5&$+0.04$&0.07\\
\multicolumn{15}{c}{BAADE's WINDOW NON-BULGE STARS} \\
II-122 &3912&59&0.14&$-0.79$&1.53&0.05&6.66&0.09& 69&6.70&0.07&4&$+0.04$&0.12\\
IV-025 &4614&66&2.64&$+0.21$&1.12&0.07&7.66&0.11& 99&7.64&0.05&4&$-0.03$&0.12\\
\multicolumn{15}{c}{LOCAL DISK STARS} \\
HR1184 &4753&11&2.27&$-0.38$&1.39&0.10&7.07&0.09&122&7.06&0.08&5&$-0.01$&0.12\\
HR1346 &4823&50&2.43&$+0.15$&1.57&0.04&7.60&0.09&113&7.59&0.03&5&$-0.01$&0.09\\
HR1348 &4409&44&1.96&$-0.36$&1.56&0.04&7.09&0.09&126&7.10&0.04&5&$+0.01$&0.10\\
HR1409 &4838&49&2.52&$+0.20$&1.63&0.04&7.65&0.09&118&7.70&0.06&5&$+0.05$&0.11\\
HR1411 &4961&63&2.69&$+0.17$&1.48&0.05&7.62&0.08&109&7.65&0.04&4&$+0.03$&0.09\\
HR1585 &4333&39&1.63&$-0.35$&1.66&0.04&7.10&0.08& 98&7.06&0.02&5&$-0.04$&0.09\\
HR2035 &4624&34&2.32&$-0.64$&1.15&0.05&6.81&0.07&122&6.82&0.07&5&$+0.01$&0.10\\
HR2113 &4277&46&1.37&$-0.64$&1.57&0.03&6.81&0.07&125&6.73&0.04&5&$-0.08$&0.08\\
HR2443 &4429&33&1.50&$-0.21$&1.61&0.03&7.24&0.07&124&7.20&0.04&5&$-0.04$&0.08\\
HR3418 &4470&70&1.91&$+0.16$&1.77&0.04&7.61&0.08&105&7.68&0.05&5&$+0.07$&0.10\\
HR3733 &5001&50&2.89&$-0.12$&1.50&0.00&7.33&0.08&121&7.40&0.07&5&$+0.07$&0.11\\
HR3905 &4531&13&2.34&$+0.32$&1.50&0.00&7.77&0.09&106&7.78&0.06&5&$+0.02$&0.11\\
HR4104 &4046&10&1.07&$-0.36$&1.87&0.05&7.09&0.10&101&7.13&0.03&5&$+0.04$&0.10\\
HR4382 &4475&72&1.72&$-0.54$&1.55&0.05&6.91&0.10&125&6.94&0.05&5&$+0.03$&0.11\\
HR4450 &4975&51&2.53&$+0.13$&1.39&0.07&7.58&0.10&111&7.55&0.10&4&$-0.03$&0.14\\
HR4608 &4853&35&2.46&$-0.41$&1.44&0.06&7.04&0.09&123&7.05&0.08&5&$+0.01$&0.12\\
HR5340 &4283&39&1.55&$-0.50$&1.61&0.03&6.95&0.07&136&6.99&0.05&5&$+0.04$&0.08\\
\enddata
\tablenotetext{a}{The value of [m/H] assumes a solar $\log{\epsilon{\rm(Fe)}} = 7.45$.  The Kurucz and Castelli grids used in this paper assume 
$\log{\epsilon{\rm(Fe)}} = 7.50$.  We use atmosphere models with 
$\log{\epsilon{\rm(Fe)}}$ values picked to match the derived Fe~I abundance.} 
\end{deluxetable}

\clearpage

\clearpage
\begin{figure}
\plotone{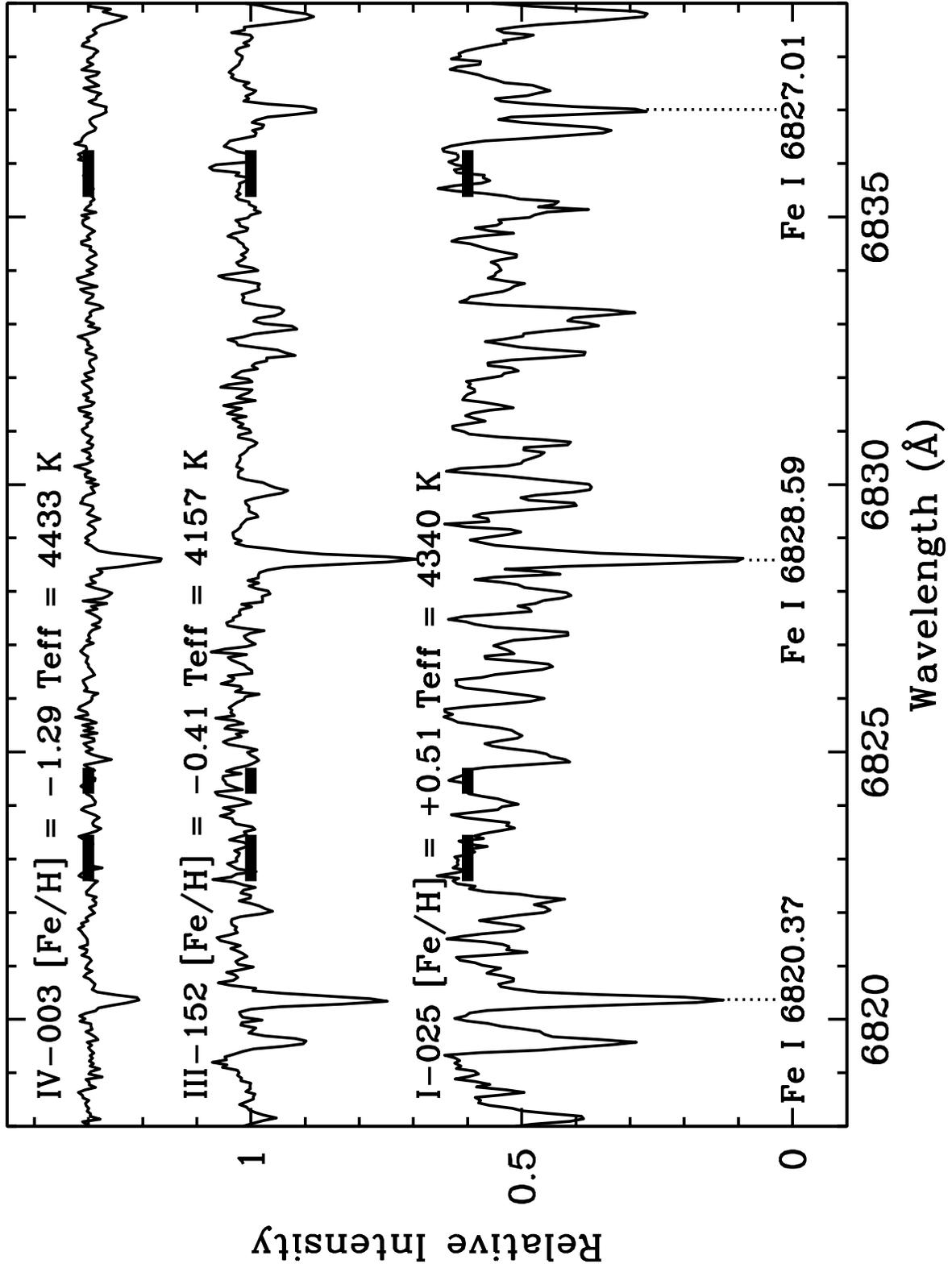}
\caption{\scriptsize Sample spectral region from three stars in the Baade's 
Window
sample.  These roughly-normalized Keck/HIRES spectra display the wide range 
of metallicities seen in the bulge.  The signal-to-noise level ranged from 
about 45 per pixel in I-025 to about 85 per pixel for III-152.  Three 
continuum regions (heavy horizontal bars with arbitrary vertical placement) and
three Fe~I lines used in the analysis are marked.  The spectrum of I-025 
demonstrates the sizable line blanketing found in very metal-rich K-giants, 
even in the red.  Most of the
additional unmarked lines in I-025 belong to CN and Fe, but measurable lines of
Si, V, Cr, La and other elements are also present in the wavelength 
interval presented.}  
\label{fig1}
\end{figure}

\begin{figure}
\plotone{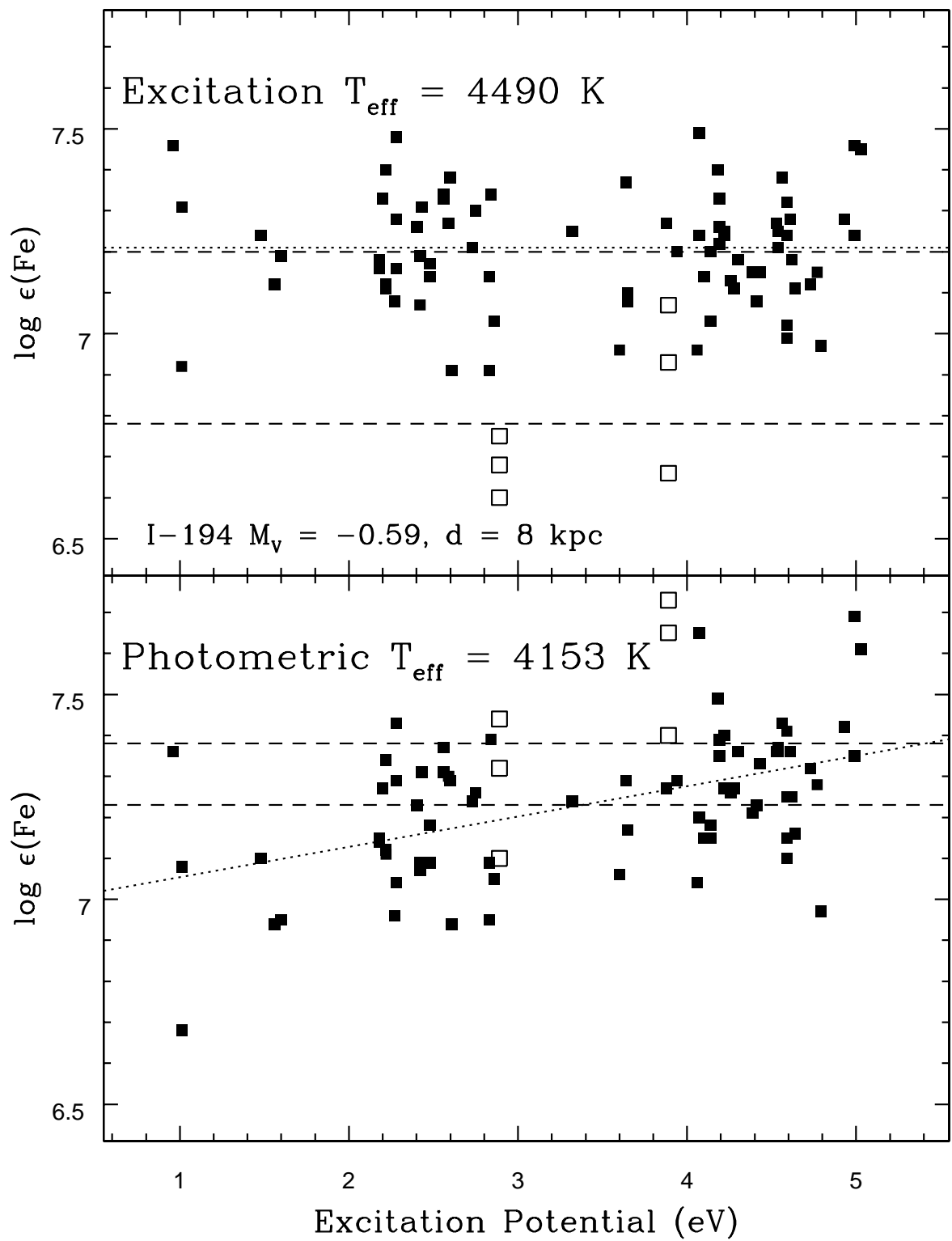}
\caption{\scriptsize Excitation plots for I-194 based on setting the 
temperature 
by the slope of the Fe~I lines (top panel) or by photometric color (lower
panel) and assuming a distance to the star of 8~kpc.  Solid points denote 
Fe~I lines, while open points mark the Fe~II
lines.  The two long-dash lines mark the mean values of \leps(Fe~I) and 
\leps(Fe~II), and the short-dash lines marks the slope of \leps(Fe~I) as 
a function of excitation potential.  In both panels, it is clear that we
cannot obtain both excitation and ionization equilibrium using the line
list and analysis method adopted for this example.  New line lists and 
analysis techniques are needed for metal-rich K-giants. \label{fig2}}
\end{figure}

\begin{figure}
\plotone{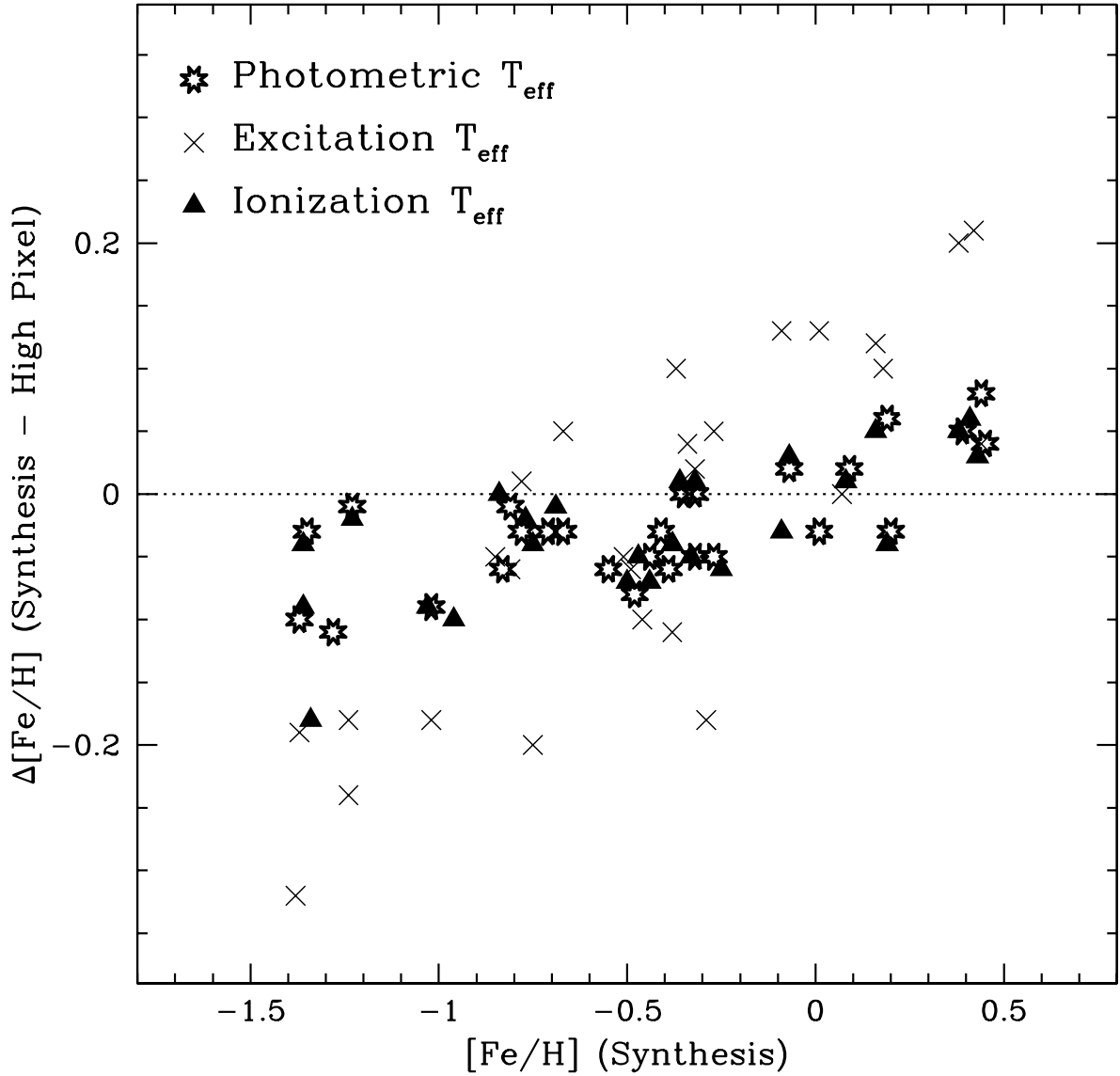}
\figcaption{\scriptsize The difference in the [Fe~I/H] abundances resulting 
from the two 
continuum setting methods for each of the 3 stellar temperature methods.
The ``High Pixel'' method exaggerates the potential systematic errors when 
setting the continuum, but is not unlike the methods used (successfully) on
weaker-lined stars.
The \tphot{} and \tion{} methods show only a slight trend with [Fe/H],
but the \tex{} method shows larger differences as a function of metallicity.
The increased slope is primarily due to the correlation between 
line excitation potential and line strength. \label{fig3}}
\end{figure}
\clearpage
\begin{figure}
\plotfiddle{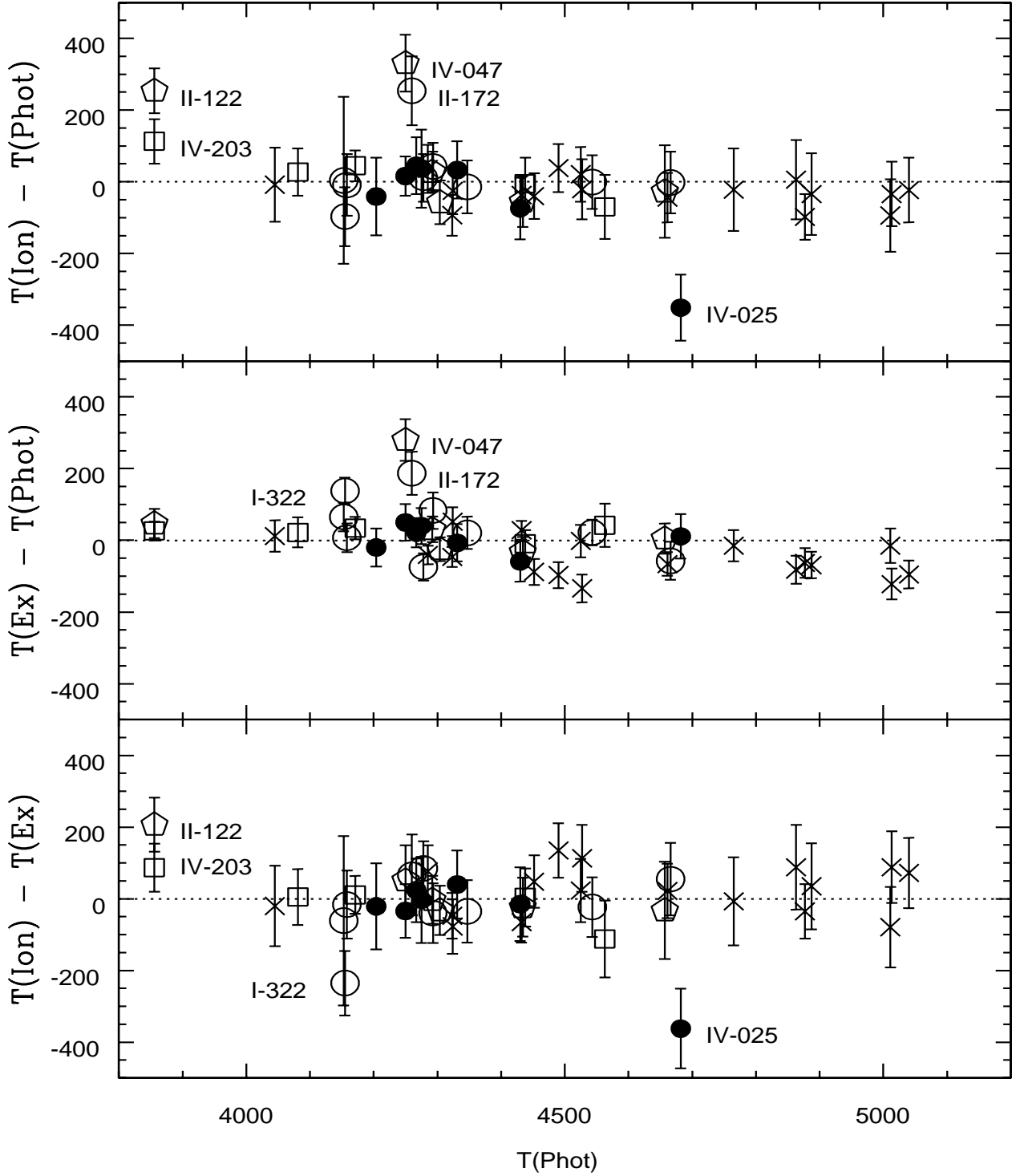}{0cm}{0.}{500.}{600.}{10}{-100}
\caption{\scriptsize Differences between the various stellar temperatures determined
by the photometric, excitation, and ionization methods for both the disk
and Baade's Window samples.  The disk giants are marked by crosses while
the Baade's Window giants are marked by symbols representing their metallicity
range:  solid circles for stars with [Fe/H] $\geq 0.0$, open circles for 
stars with $-0.50 \leq$ [Fe/H] $< 0.0$, open pentagons for stars with 
$-1.0 \leq$ [Fe/H] $< -0.5$, and open squares for stars with [Fe/H] $< -1.0$.
The surface gravities used for these 
determinations were set by assuming a distance of 8~kpc for the Baade's
Window sample.  The disk stars and most of the Baade's Window sample show 
excellent agreement between all three methods.  The success of the disk giant
analysis helps show that our methods work well when well-determined
inputs (distance, reddening, etc.) are applied.  The six exceptions from the
Baade's Window sample that are discussed in the text are labeled. \label{fig4}}
\end{figure}

\begin{figure}
\plotone{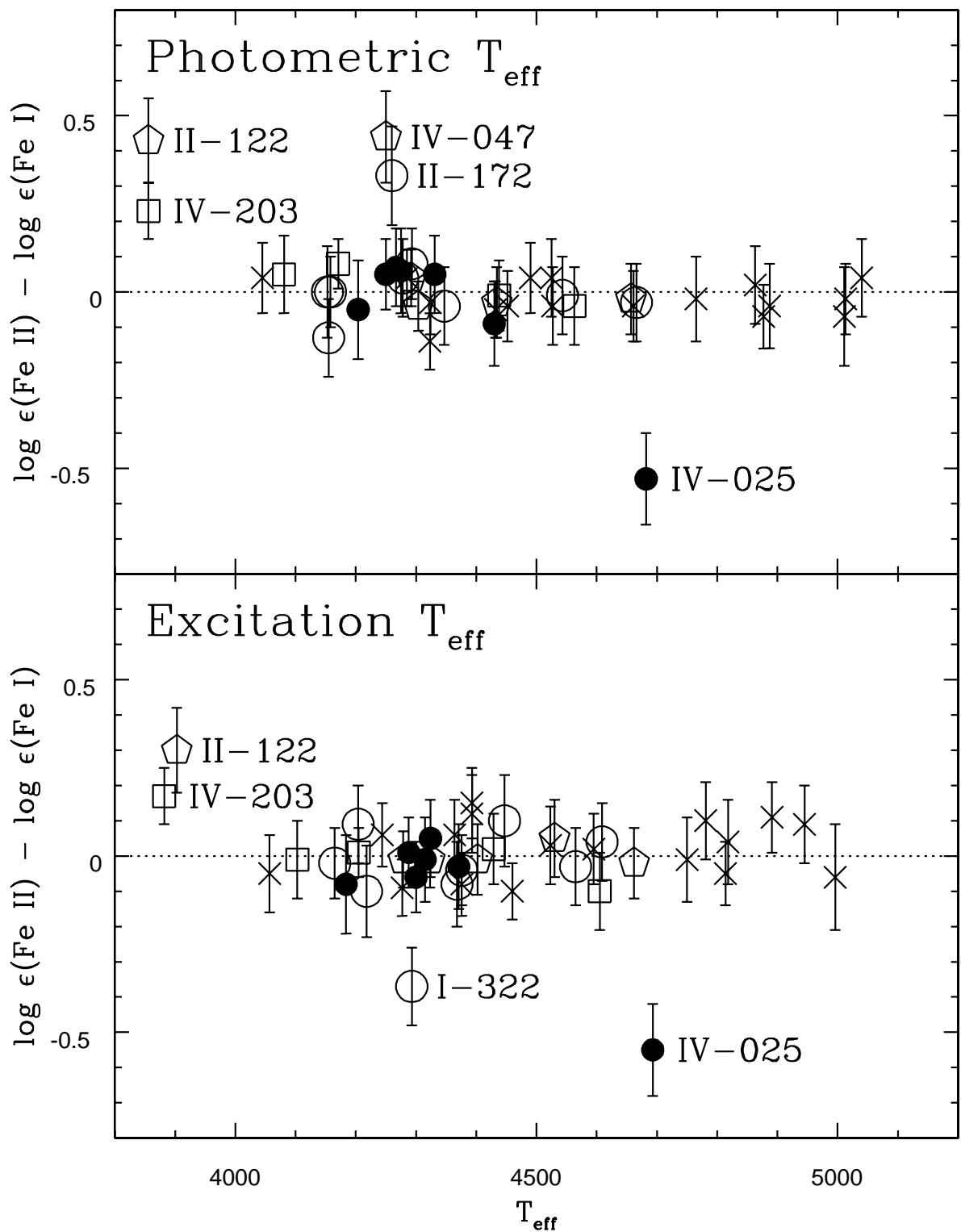}
\caption{\scriptsize Ionization differences plotted as a function of \teff{} for
the photometric and excitation temperatures methods.  The symbols are the
same as those used in Figure 4.  The surface gravities 
used for these determinations were set by assuming a distance of 8~kpc for 
the Baade's Window sample.  The disk stars and most of the Baade's Window 
sample show small ionization differences independent of method.  
The six exceptions from the
Baade's Window sample that are discussed in the text are marked. \label{fig5}}
\end{figure}

\begin{figure}
\plottwo{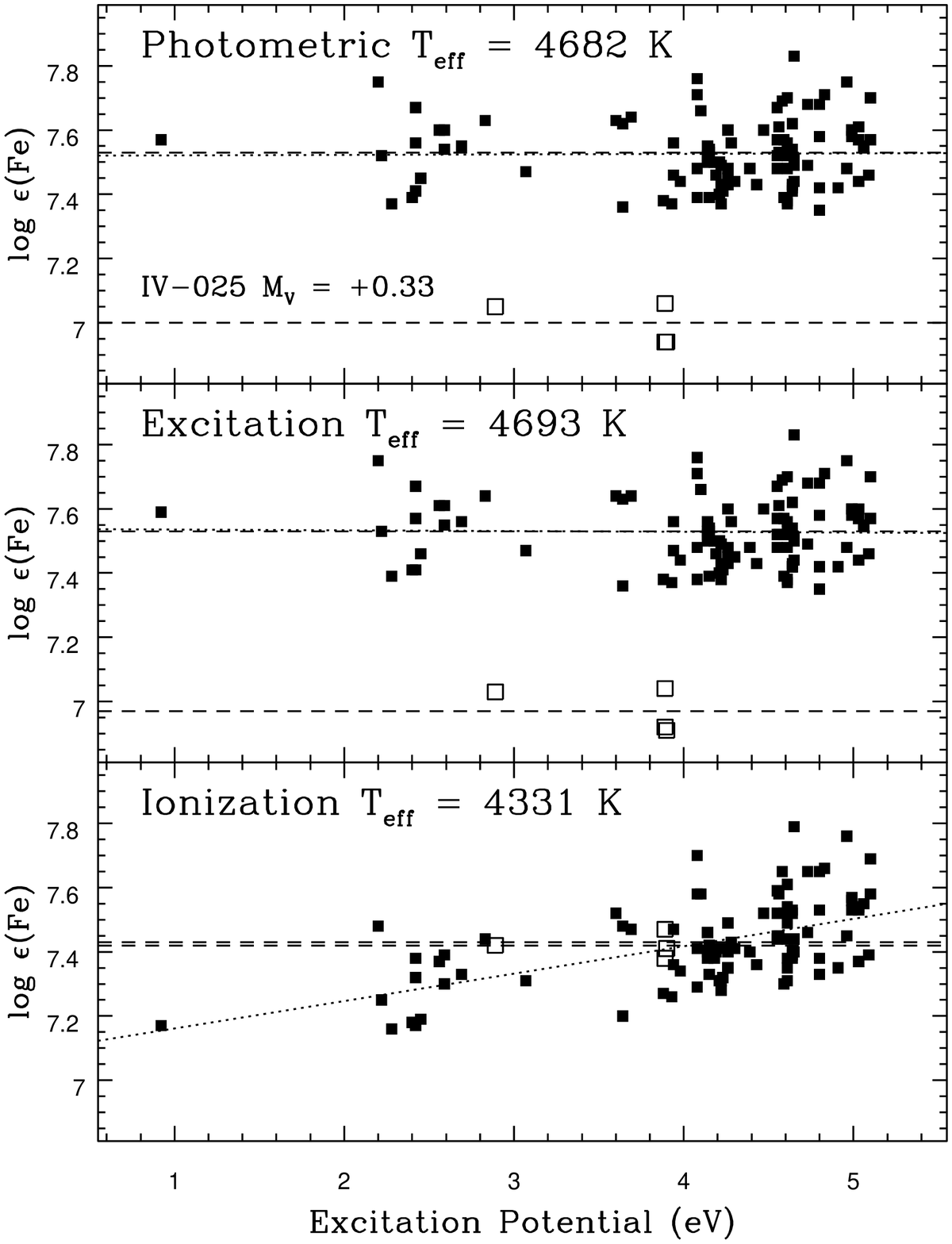}{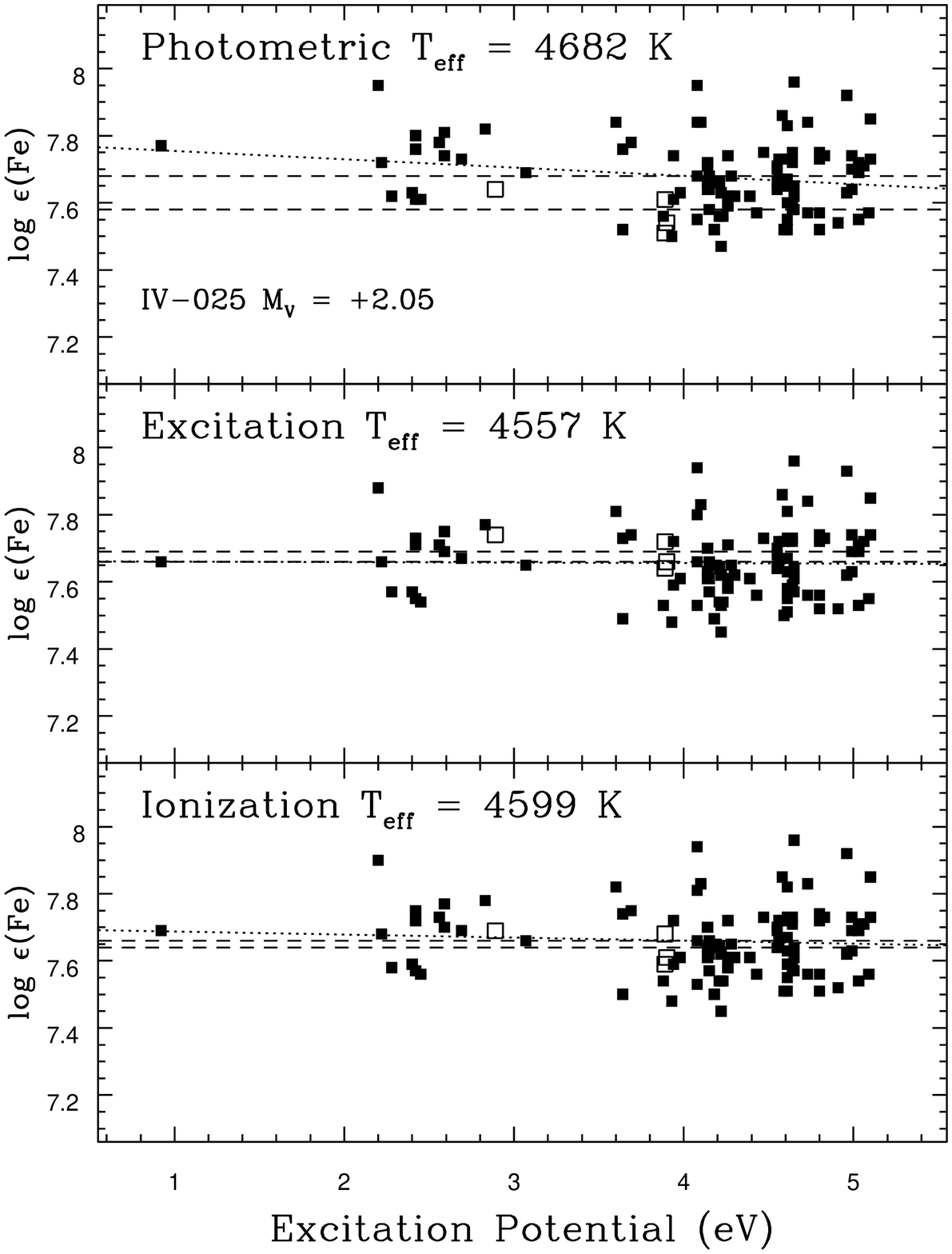}
\figcaption{\scriptsize Excitation plots for the three different 
temperature methods
for the Baade's Window star IV-025.  The solid points mark individual Fe~I
lines, while the open points mark Fe~II lines.  The two long-dash lines
mark the mean values of \leps(Fe~I) and \leps(Fe~II), and the short-dash
lines marks the slope of \leps(Fe~I) as a function of excitation potential.
The left panel shows the plots for an assumed distance modulus of 14.51~mag
(8.0~kpc).  The ionization difference for the \tphot{} and \tex{} methods
are $-0.53$ and $-0.55$~dex, respectively.  The right panel shows the same
plots when the distance modulus is changed to 12.79~mag (3.6~kpc).  This 
distance minimizes the ionization and \teff{} differences.  \label{fig6}}
\end{figure}

\begin{figure}
\plotone{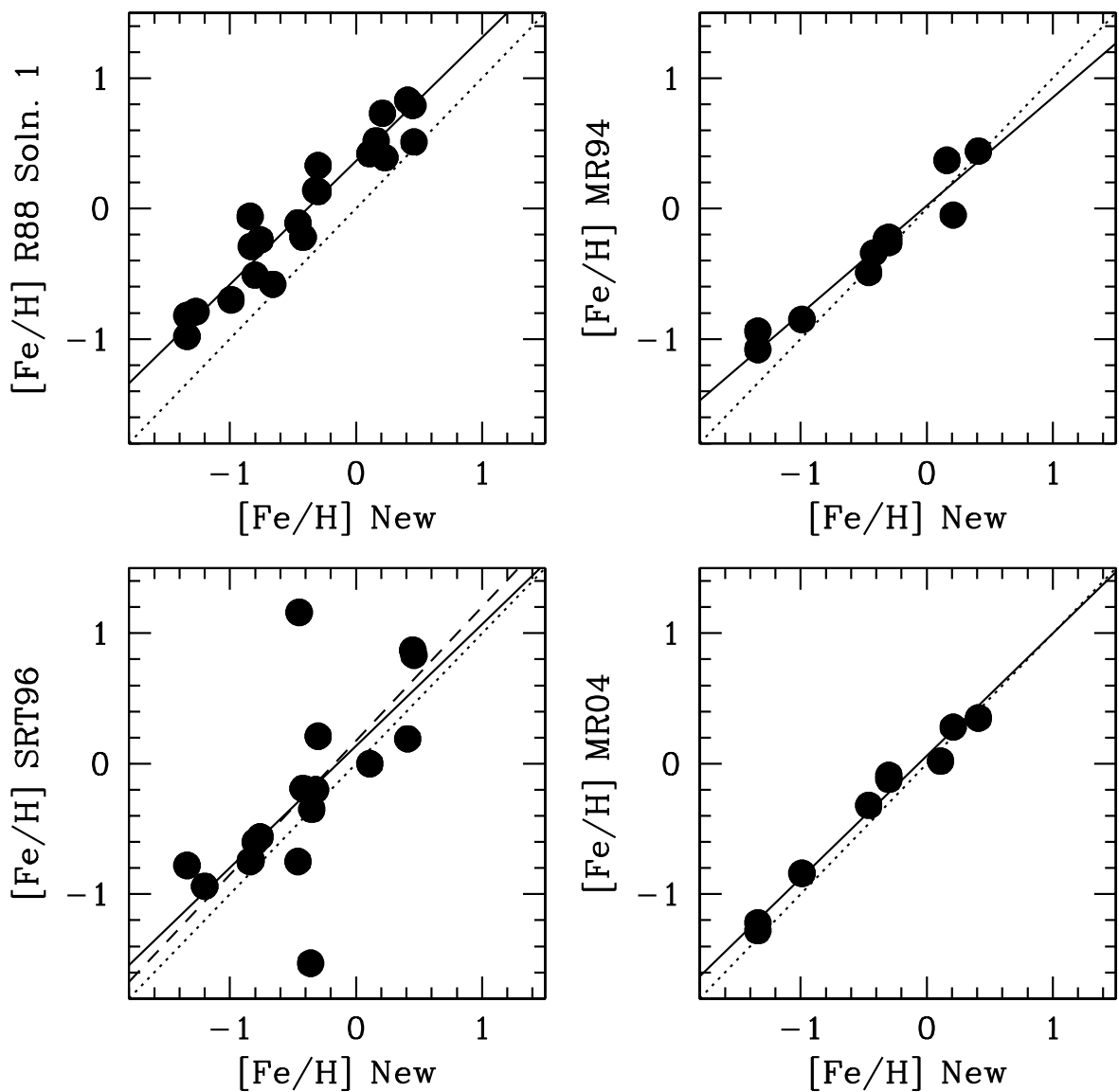}
\caption{\scriptsize Comparison of the [Fe/H] values determined in this work
to previous studies of bulge stars (Rich~1998~=~R98, 
McWilliam~\&~Rich~1994~=~MW94, Sadler~et~al.~1996~=~SRT96, and 
McWilliam~\&~Rich~2003~=~MR04).  The dotted lines are the one-to-one
line, and the solid lines are the least-squares fits given in Section~7.3. 
For the fits to the SRT96 data, the dashed line is the fit to all 17 stars
(Equation~6), while the solid line is the fit when the two outlier stars
are removed from the sample (Equation~10).
\label{fig7}}
\end{figure}

\begin{figure}
\plotone{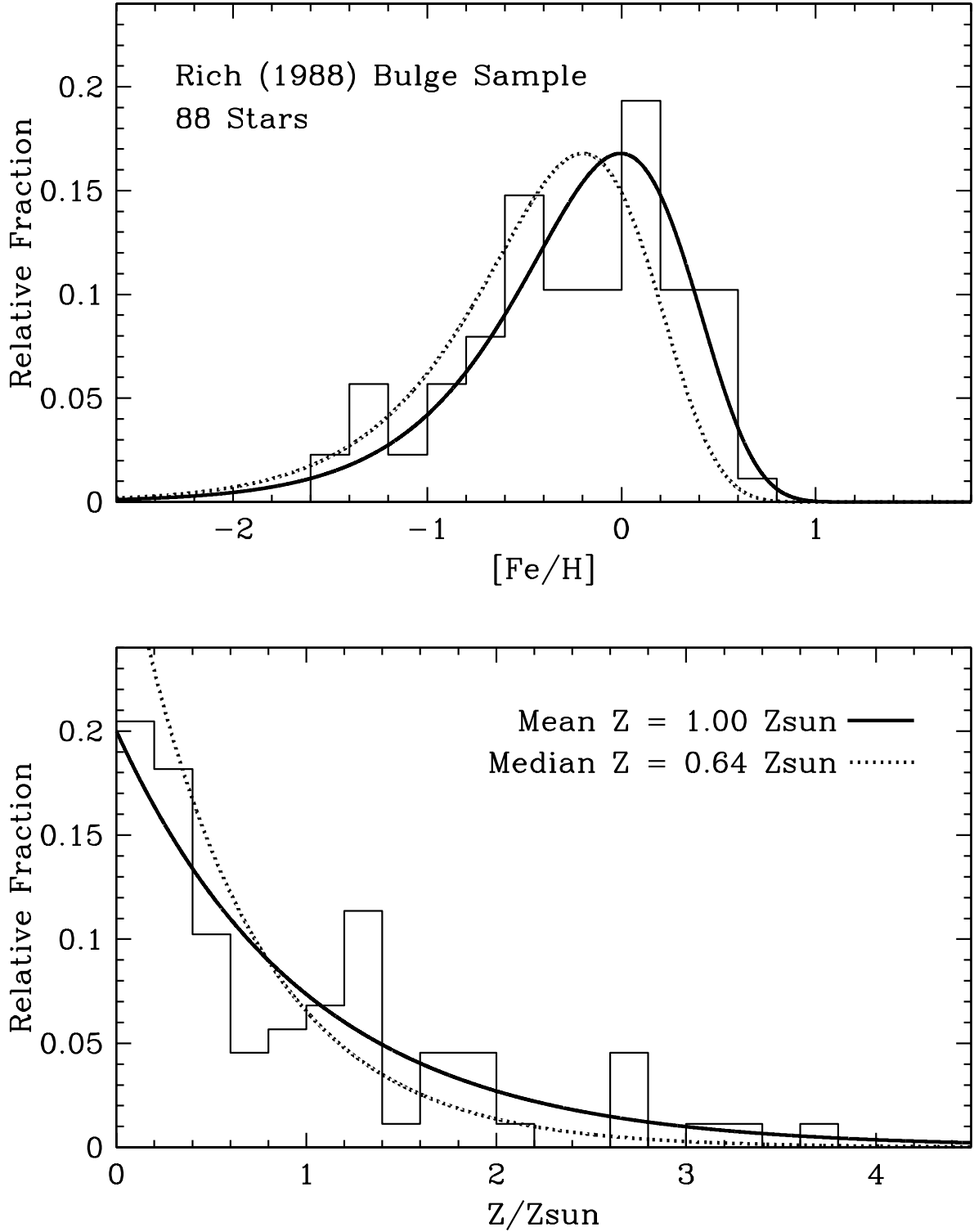}
\figcaption{\scriptsize Histograms of the [Fe/H] value (top) and heavy metal mass fraction 
values~(Z) (bottom)
for the recalibrated data of Rich~1988.  The recalibration assumes
solar abundance ratios for all the heavy elements.  Previous high-resolution
studies have found that bulge stars have enhanced [X/Fe] ratios for several
elements, so the values of~Z are probably lower limits.  The two lines are 
the closed box gas exhaustion model where the value of the yield is either
the mean or median Z~value.   \label{fig8}}
\end{figure}

\begin{figure}
\plotone{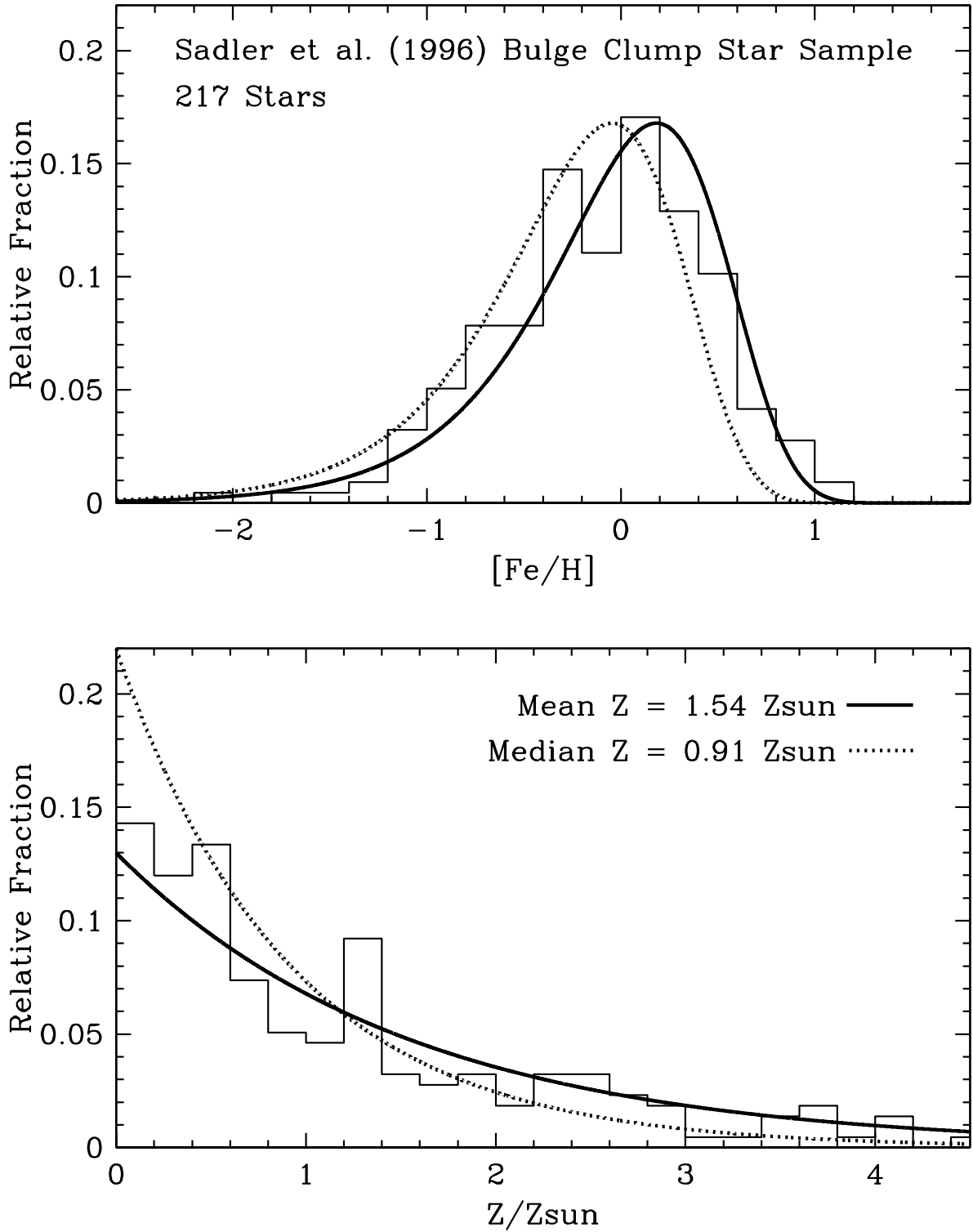}
\caption{\scriptsize Same as Figure 8, except for the recalibrated RGB clump star data 
of Sadler et al.~1996.  The recalibrated data shows more metal-rich stars 
than the recalibrated R88 sample.  There is a wider difference between
the mean and median values of Z, and the best yield for the closed box
model most likely falls between the two.  \label{fig9}}
\end{figure}

\begin{figure}
\plotone{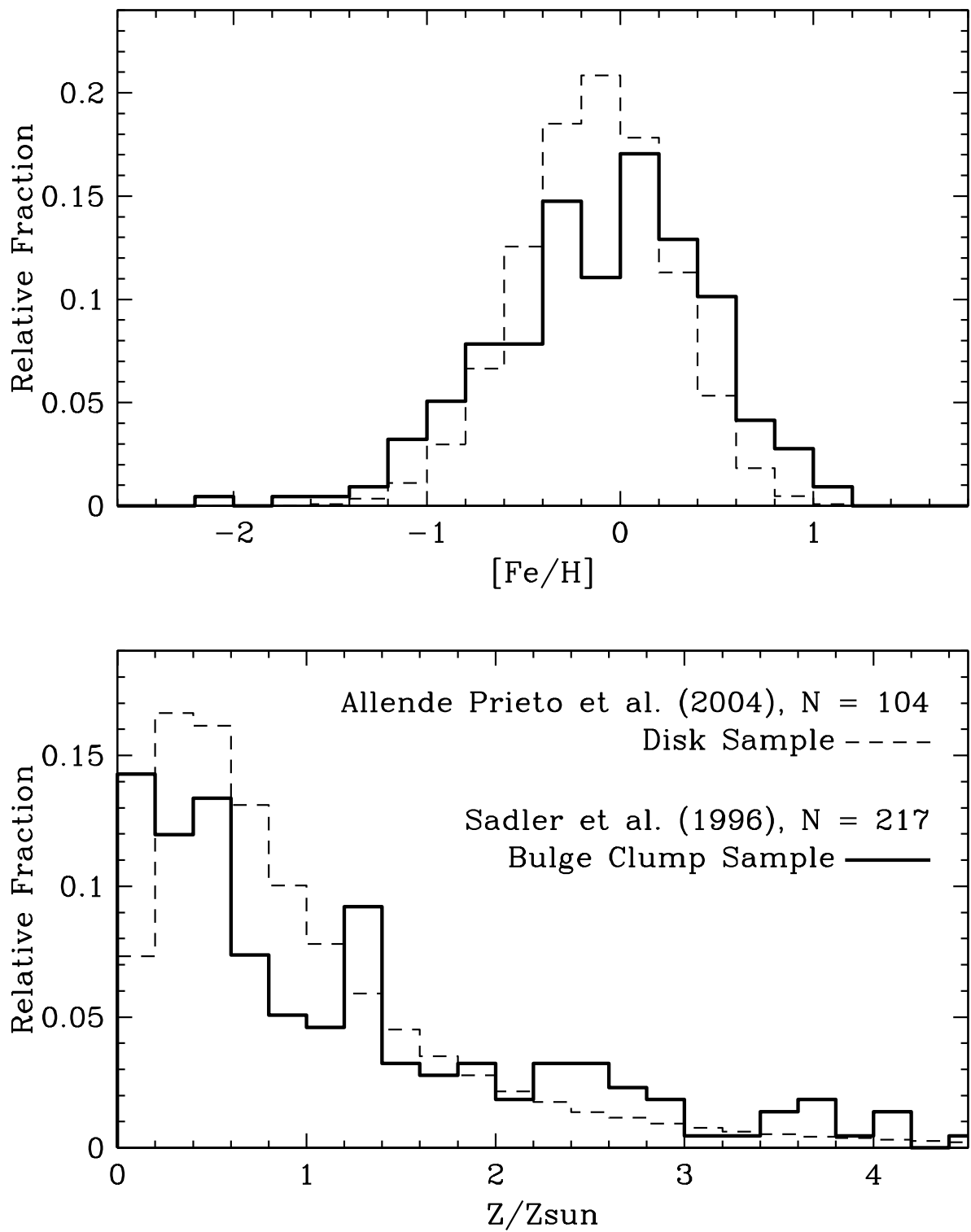}
\caption{\scriptsize Histogram of [Fe/H] values (top) and Z values (bottom)
for the recalibrated RGB clump star data of Sadler et al.~1996 and the solar 
neighborhood
sample of Allende Prieto et al.~2004.  The disk sample has been convolved
with a Gaussian with $\sigma = 0.283$~dex in order to degrade the quality
of the abundance determinations to the same quality as the SRT96 recalibration.
In the top panel, the bulge sample shows a slightly higher mean Fe abundance 
and a 
wider spread in metallicity.  In the bottom panel, the disk sample shows
a drop in the relative fraction in the most metal-poor bin, but an exponential
tail to higher metallicities.  The bulge sample does not show the same lack of 
metal-poor stars as seen in the disk sample.  \label{fig10}}
\end{figure}

\end{document}